\documentclass[onecolumn]{emulateapj}
\usepackage{amsmath,amssymb,verbatim,graphicx,rotating}
\usepackage{psfrag}

\usepackage{bigdelim,bigstrut,multirow}
\newcommand{\minitab}[2][l]{\begin{tabular}{#1}#2\end{tabular}}

\usepackage{color}
\definecolor{red}           {cmyk}{0,1,1,0}
\newcommand\jcap{{J. Cosmology Astropart. Phys.}}
\newcommand{\be}{\begin{equation}}
\newcommand{\ee}{\end{equation}}
\newcommand{\bea}{\begin{eqnarray}}
\newcommand{\eea}{\end{eqnarray}}

\newcommand{\avg}[1]{\langle #1 \rangle}

\shorttitle{{\it Planck} precision on $r$ \& other parameters}
\shortauthors{Burigana et al.}

\begin{document}


\title{Forecast for the {\it Planck} precision on the tensor to scalar ratio 
and other cosmological parameters}


\author{C. Burigana\altaffilmark{1}, C. Destri\altaffilmark{2}, H. J. de Vega\altaffilmark{3,4},
A. Gruppuso\altaffilmark{1},}
\author{N. Mandolesi\altaffilmark{1}, P. Natoli\altaffilmark{5},  N. G. Sanchez \altaffilmark{4}}

\altaffiltext{1}{INAF/IASF, Istituto di Astrofisica Spaziale e 
Fisica Cosmica di Bologna,\\
Istituto Nazionale di Astrofisica, via Gobetti 101, I-40129 Bologna, Italy.}
\altaffiltext{2}{Dipartimento di Fisica G. Occhialini, Universit\`a
Milano-Bicocca and INFN,\\ sezione di Milano-Bicocca, Piazza della Scienza 3,
20126 Milano, Italy.}
\altaffiltext{3}{LPTHE, Laboratoire Associ\'e au CNRS UMR 7589,\\
Universit\'e Pierre et Marie Curie (Paris VI) et Denis Diderot 
(Paris VII), \\ Tour 24, 5 \`eme. \'etage, 4, Place Jussieu, 
75252 Paris, Cedex 05, France.}
\altaffiltext{4}{Observatoire de Paris, LERMA, Laboratoire Associ\'e au 
CNRS UMR 8112, \\61, Avenue de l'Observatoire, 75014 Paris, France.}
\altaffiltext{5}{Dipartimento di Fisica, Universit\`a di Roma Tor Vergata and INFN,\\ sezione di Tor Vergata, 
Via della Ricerca Scientifica 1, I-00133 Roma, Italy.}


\begin{abstract}
The {\it Planck} satellite, successfully launched on May 14th 2009 to measure with unprecedented 
accuracy the primary Cosmic Microwave Background (CMB) anisotropies, is 
operating as expected. The Standard Model of the Universe 
(``concordance'' model) provides the current realistic context to analyze the 
CMB and other cosmological/astrophysical data, inflation in the early Universe being 
part of it. The {\it Planck} performance for the crucial 
primordial parameter $r$, the tensor--to--scalar ratio related to primordial $B$ mode polarization,
will depend on the quality of data analysis and interpretation.
The Ginzburg-Landau approach to inflation allows 
to take high benefit of the CMB data. The fourth degree double well inflaton potential gives an 
excellent fit to the current CMB+LSS data. We evaluate the {\it Planck} precision to the recovery of
cosmological parameters, taking into account a reasonable toy model for residuals of
systematic effects of instrumental and astrophysical origin based on publicly available information. 
We use and test two relevant models: the $\Lambda$CDM$r$ model, 
i.e. the standard $\Lambda$CDM model augmented by $ r $, and the $\Lambda$CDM$r$T model, 
where  the scalar spectral index,  $n_s$, and $r$ are  related through the theoretical 
``banana-shaped'' curve $ r = r(n_s)$ coming from the 
Ginzburg-Landau theory with double--well inflaton potential.  
In the latter case,  the analytical expressions for $ n_s $ and $ r $ are 
imposed as a hard constraint in the Monte Carlo Markov Chain (MCMC) data 
analysis.
 
We consider two $C_\ell-$likelihoods (with and without $B$ 
modes) and take into account the white noise
sensitivity of {\it Planck} (LFI and HFI) in the 70, 100 and 143 GHz channels 
as well as the residuals from systematics errors and foregrounds. 
We also consider a cumulative channel of the three ones.  We produce the sky (mock data) 
for the CMB multipoles $ C_l^{TT},  \; C_l^{TE}, \; C_l^{EE} $ and $ C_l^{BB} $ from the  
$\Lambda$CDM$r$ and  $\Lambda$CDM$r$T models and obtain the cosmological parameter marginalized 
likelihood distributions for the two models. Foreground residuals turn to affect only the 
cosmological parameters sensitive to the $ B $ modes. As expected, the likelihood $ r $ distribution is much
clearly peaked near the fiducial value ($ r = 0.0427 $) in the $\Lambda$CDM$r$T model than 
in the $\Lambda$CDM$r$ model. The best value for $ r $ in the presence of residuals
turns to be about $ r \simeq 0.04 $ for both the $\Lambda$CDM$r$ and the 
$\Lambda$CDM$r$T models. The $\Lambda$CDM$r$T model turns to be very stable, 
its distributions do not change by including residuals and the $ B $ modes. 
For $ r $ we find  $ 0.028 < r < 0.116 $   at 95 \%  CL  with the best value $ r = 0.04 $.
We also compute the $B$ mode detection probability by the most sensitive HFI-143 channel.
At the level of foreground residual equal to 30\% of our toy model only a 68\% CL (one sigma)
detection is very likely. For a 95\% CL detection (two sigmas)
the level of foreground residual should be reduced to 10\% or lower of the adopted toy model.
The lower bounds (and most probable value) we
infer for $ r $ support the searching of CMB $B$ mode polarization  
in the current data as well as the planned CMB missions oriented to $B$ polarization.
\end{abstract}





\section{INTRODUCTION AND WORK OUTLINE} 

The {\it Planck} satellite\footnote{http://www.rssd.esa.int/planck}  was successfully launched on May 14th 2009 to 
measure the primary Cosmic Microwave Background (CMB)
temperature and polarization anisotropies on the whole sky with unprecedented accuracy. 
It is now in normal operation, with the expected performances 
\citep{PlanckBlueBook,prelauB,prelauM,lamarre2010,Maffei_prelaunch}. {\it Planck}  will 
improve the measurement 
of most cosmological parameters by several factors 
with respect to current experiments, 
in particular the Wilkinson Microwave Anisotropy Probe (WMAP) satellite\footnote{http://lambda.gsfc.nasa.gov/}.
The expected CMB polarization measurements from {\it Planck} will allow to push both
($E$ and $B$) polarization results well beyond the present knowledge and considerably constrain 
the tensor ($B$ modes) to scalar ratio parameter $ r $, if not to obtain a detection on it. 
In this respect, the way of extracting and physically interpreting cosmological parameters 
(once the CMB data cleaned from the different astrophysical foregrounds) 
will be important.
In other words, the {\it Planck} actual performance 
for the crucial primordial parameter $ r $ will depend on 
the adopted physical modeling and on the quality of data analysis and interpretation.
It is then important and timely to make 
forecasts for the {\it Planck} determination of $ r $ and other cosmological parameters
taking into account the theoretical progress in the field and WMAP results.

\medskip

The Standard Model of the Universe (or ``concordance'' model) provides the current 
realistic context to analyze the CMB and other cosmological/astrophysical 
data. Inflation (quasi-exponential accelerated expansion) of the early Universe 
is a part of this model and one important goal of CMB experiments is probing the 
physics of it. Inflation solves the shortcomings of the decelerated expanding cosmology 
(horizon problem, flatness, entropy of the Universe), and explains the observed CMB 
anisotropies providing the mechanism for the generation of scalar and tensor perturbations 
seeding the large scale structures (LSS) and primordial (still undetected) gravitational 
waves ($B$ mode polarization).

\medskip

The current CMB $+$ LSS data support the standard inflationary 
predictions of a nearly spatially flat Universe with adiabatic 
and nearly scale invariant 
initial density perturbations. These data  are validating the single field slow-roll 
inflationary scenario \citep{WMAP5}. Single field slow-roll models provide an appealing, 
simple and fairly generic description of inflation \citep{libros,reviu}. The 
inflationary scenario is implemented using a scalar field, the \emph{inflaton} with a 
potential $ V(\varphi) $, self-consistently coupled to the space-time metric. 
In the effective theory based on the Ginzburg-Landau (G-L) approach to inflation \citep{reviu}, 
the potential is a polynomial in the field starting by a constant term. Linear terms can always 
be eliminated by a constant shift of the inflaton field. 
The mass (quadratic) term can have a positive or a negative sign 
associated to unbroken symmetry (chaotic inflation) or to broken symmetry
(new inflation), respectively. The fourth degree double--well 
inflaton potential gives an excellent fit of the present CMB $+$ LSS data \citep{reviu}.
A cubic term does not improve the fit and can be omitted \citep{mcmc1}.
Adding higher order terms with additional parameters does not improve significantly
the fits \citep{high}. The G-L framework is not just a class of physically well
motivated inflaton potentials, among them the double and single well
potentials. This approach provides the effective theory  for
inflation, with powerful gain in the physical insight and analysis of the
data. The present set of data with the effective theory of inflation favor
the double well potential  \citep{reviu,mcmc1}. 
Analyzing the present data
without the relation between $r$ and $n_s$  does not allow
to discriminate among different classes of models for the inflaton potential
in the considered framework.
Although the G-L effective theory approach to inflation is quite general,
it predicts precise
order of magnitude estimates for $ n_s $, $ r $ and the running of the spectral 
index $ dn_s/d \ln k $ \citep{reviu}
$$
n_s - 1 = {\cal O}\left(\frac1{N}\right) \; , \quad 
r = {\cal O}\left(\frac1{N}\right) \; , \quad
\frac{d n_s}{d \ln k} =  {\cal O}\left(\frac1{N^2}\right) \; ;
$$
here  $ N \sim 60 $ is the number 
of efolds since the cosmologically relevant modes exit the horizon till 
inflation ends. The WMAP values for $ n_s $ and the upper bounds for $ r $ and 
 $ dn_s/d \ln k $ agree with these estimates. Since in this framework the estimated running,
$ dn_s/d \ln k \sim 3 \times 10^{-4} $, is very small,  
in this paper we will concentrate on $ n_s $ and $ r $.

\medskip

In this work, we evaluate the accuracy in the recovery of the cosmological parameters 
expected from the {\it Planck} data. First, we do this forecast without including the systematic
effects of instrumental and/or astrophysical origin or
their coupling, affecting the {\it Planck} measurements, and then by including 
the systematic effects. 
In this study we exploit the {\it Planck} sensitivity and resolution at its three 
favorite cosmological 
channels, i.e. at the frequencies of 70, 100, and 143 GHz. Table \ref{table:sens} 
reports the {\it Planck} performance at these frequencies, based on 
\citet{PlanckBlueBook} and, for the LFI channel at 70 GHz, as updated in 
\citet{prelauM}, \citet{prelauB}, \citet{sandri_etal_2010}.
%HFI performance at 100 and 143 GHz is in agreement with \citep{PlanckBlueBook}.
These sensitivities do not include the degradation in accuracy that could come from 
various sources of systematic effects, of both instrumental and/or astrophysical origin, or
their coupling.
In Sect. \ref{toym} we discuss the current published estimates for the residuals 
of systematic effects 
and foregrounds affecting the {\it Planck} CMB measurements: straylight, main beam asymmetry, 
leakage, time constants, glitches, and foregrounds. 
In general, we do not use in this  work a precise (still not completely available) description
of the considered systematic effects, but only suitable representations of them, as described 
in Sect. \ref{toym}. This is done in a parametric approach, identifying the corresponding levels at 
which the control of the systematic effects is necessary not to spoil 
the {\it Planck} data scientific accuracy. 
We technically implement this rescaling with a multiplicative constant on the 
residuals of the systematic effects on the CMB multipoles $ C_\ell $.
Obviously, the real analysis of {\it Planck} data will have to properly consider 
all possible systematic effects of optical, thermal, and instrumental 
(radiometric and bolometric) origin, with an even better accuracy than
those achieved in past projects. In parallel, a significantly 
improved separation of CMB from astrophysical components
will be needed, a task in principle possible for {\it Planck}  
thanks to its wide frequency coverage.

\medskip

Instrumental systematics on CMB tensors-to-scalar have been studied by
\citet{whu,shimon,yadav}. 

\medskip

We use and test two relevant models: the 
$\Lambda$CDM$r$ model, that is the standard $\Lambda$CDM model augmented by
the tensor--to--scalar ratio $ r $, and the $\Lambda$CDM$r$T model, 
that is the $\Lambda$CDM$r$ model in which
the double--well inflaton potential (see Eq. (\ref{binon}) in the next section) is imposed. Namely,
$ n_s $ and $ r $ are constrained by the analytic relation $ r = r(n_s) $
to lay on the theoretical banana-shaped curve (the upper 
border of the banana-shaped region Fig. \ref{banana}). 
The novelty in the MCMC analysis of the CMB data with the $\Lambda$CDM$r$T model
is in the fact that we impose the analytical expressions for $
  n_s $ and $ r $ derived from the inflaton potential as a hard
constraint \citep{mcmc1}. We take both models, $\Lambda$CDM$r$ and $\Lambda$CDM$r$T, 
as fiducial models in our Monte Carlo Markov Chains (MCMC) simulations
to produce the corresponding skies (mock data). 
In the $\Lambda$CDM$r$ model the independent cosmological
parameters are $\Omega_b \, h^2, \; \Omega_c \, h^2, \; \theta , \; \tau, \; A_s ,
\; n_s $ and $ r $, while all other independent parameters are assumed to vanish,
{\em e.g.} $ \Omega_\nu=0 $, or have the standard values, {\em e.g.}  $ w=-1 $.
The aforementioned $\Lambda$CDM$r$T model includes the same parameters but with $ n_s $ and $ r $ not
being independent, but related by the curve $r=r(n_s)$ as widely discussed in Sect. \ref{LGtheory}. 
We produce one sky (mock data) for the anisotropy CMB multipoles $ C_l^{TT}, \; C_l^{TE}, \;
C_l^{EE} $ and $ C_l^{BB} $ from the  $\Lambda$CDM$r$ model and from the $\Lambda$CDM$r$T
model, with the parameters in Table \ref{tab2}. We describe the detailed procedure in Sect. \ref{mock}. 
We run Monte Carlo Markov Chains from this sky and obtain the marginalized
  likelihood distributions for the cosmological parameters ($ \Omega_b \, h^2 , \;
    \Omega_c \, h^2 ,\; \theta \; \tau, \; \Omega_{\Lambda} $, Age of the
    Universe, $ z_{re}, \; H_0, \; A_s, \;  n_s $ and $ r $) in the two test models
    $\Lambda$CDM$r$ and $\Lambda$CDM$r$T . 
    We study the independent $\Lambda$CDM$r$ parameters with the mock
    data produced from $\Lambda$CDM (first row of Table \ref{tab2}) and the
    independent parameters of both $\Lambda$CDM$r$ and $\Lambda$CDM$r$T with the
    mock data produced from $\Lambda$CDM$r$T (second row in Table \ref{tab2}). 
The fiducial values, $ r = 0.0427 $ and $ n_s = 0.9614 $ correspond to the best fit to the 
CMB-LSS data with the $\Lambda$CDM$r$T model using the double--well inflaton 
potential expressed by Eq. (\ref{binon}). 

Namely, these are the best fit values to
$ r $ and $ n_s $ within the Ginsburg-Landau effective theory approach. Not 
using the Ginsburg-Landau approach, lower bounds for $ r $ are not obtained
and the best fit value for $ r $ can be much smaller than $ r = 0.04 $ \citep{kinney08,PeirisEasther}.

We consider two choices for the $C_\ell-$likelihood, 
one without the $ B $ modes and one with the $B$ modes and take into account the white noise
sensitivity of {\it Planck} (LFI and HFI) in the 70, 100 and 143 GHz channels
\citep{PlanckBlueBook}. We also consider a cumulative channel whose $ \chi^2 $ is the
  sum of the $ \chi^2 $'s of the three channels above. When using different
  channels in the MCMC analysis, we use different noise realizations while
  keeping the same sky, that is the same realization of the Gaussian
  process that generated the primordial fluctuations. 
In our MCMC analysis we always take standard flat priors for the cosmological
  parameters. In particular we assume the flat priors $ 0 \leq r < 0.2 $ in the
  $\Lambda$CDM$r$ model and $ 0 \leq r< 8/60$, where $8/60 \simeq 0.133$ 
  is the theoretical upper limit for $ r $   in the $\Lambda$CDM$r$T model.
 
We performed the MCMC simulations using the publicly available CosmoMC 
code\footnote{http://cosmologist.info/cosmomc/} \citep{mcmc} interfaced to the 
Boltzmann code CAMB\footnote{http://camb.info/} (see \citet{2000ApJ...538..473L} and references therein).
\medskip 

\begin{figure}
  \includegraphics[height=7cm]{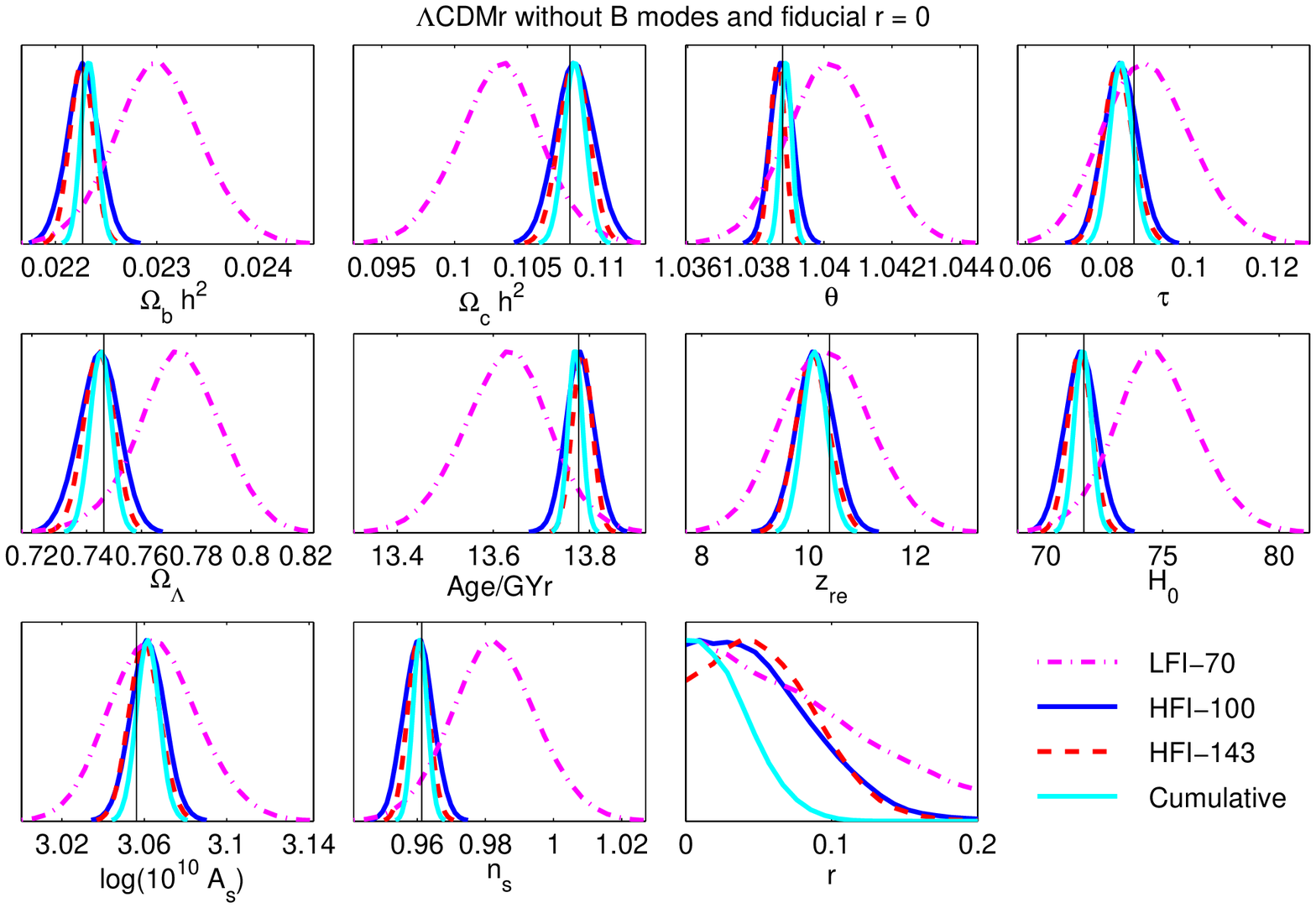} \\
  \includegraphics[height=7cm]{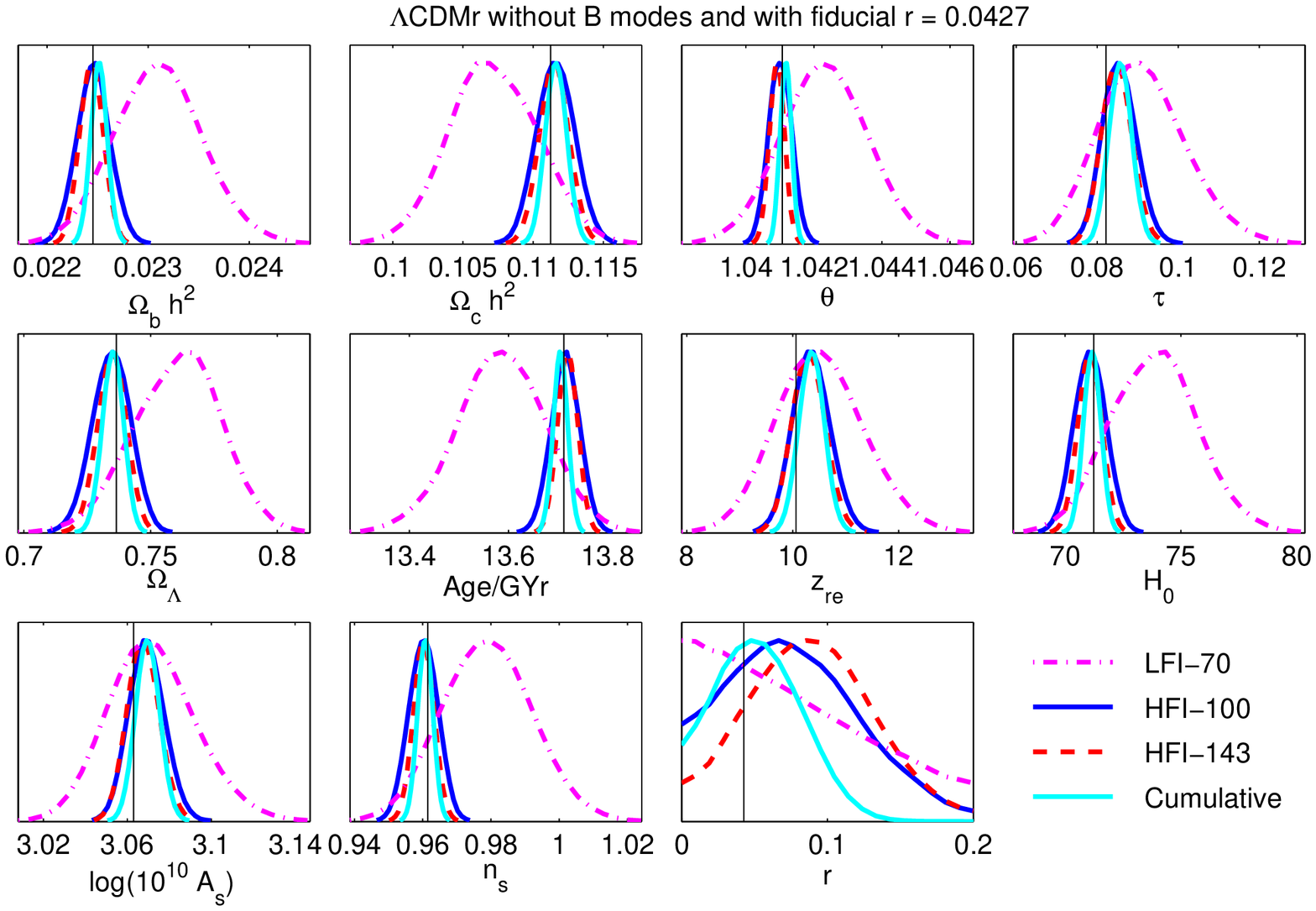}
  \caption{Marginalized likelihood distributions, without including $ B $ modes,
    of the cosmological parameters for the $\Lambda$CDM$r$ model.
We display the distributions for each of the three channels  HFI-100, HFI-143 and LFI-70
and for the cumulative of the three channels. 
The fiducial values are indicated by a vertical thin black line. 
The fiducial value for the ratio is $ r = 0 $ in the upper panel and $ r =  0.0427 $ in
the lower panel. Notice that the latter fiducial value is smaller than the peaks of
    the marginalized distribution. This is due just to statistical fluctuations
since we are considering only one sky. The upper panel figures imply the upper bound 
$ r < 0.068 $ at $ 95\%$ CL without $ B $ modes.}
  \label{r0r04}
\end{figure}

\begin{figure}
  \includegraphics[height=7cm]{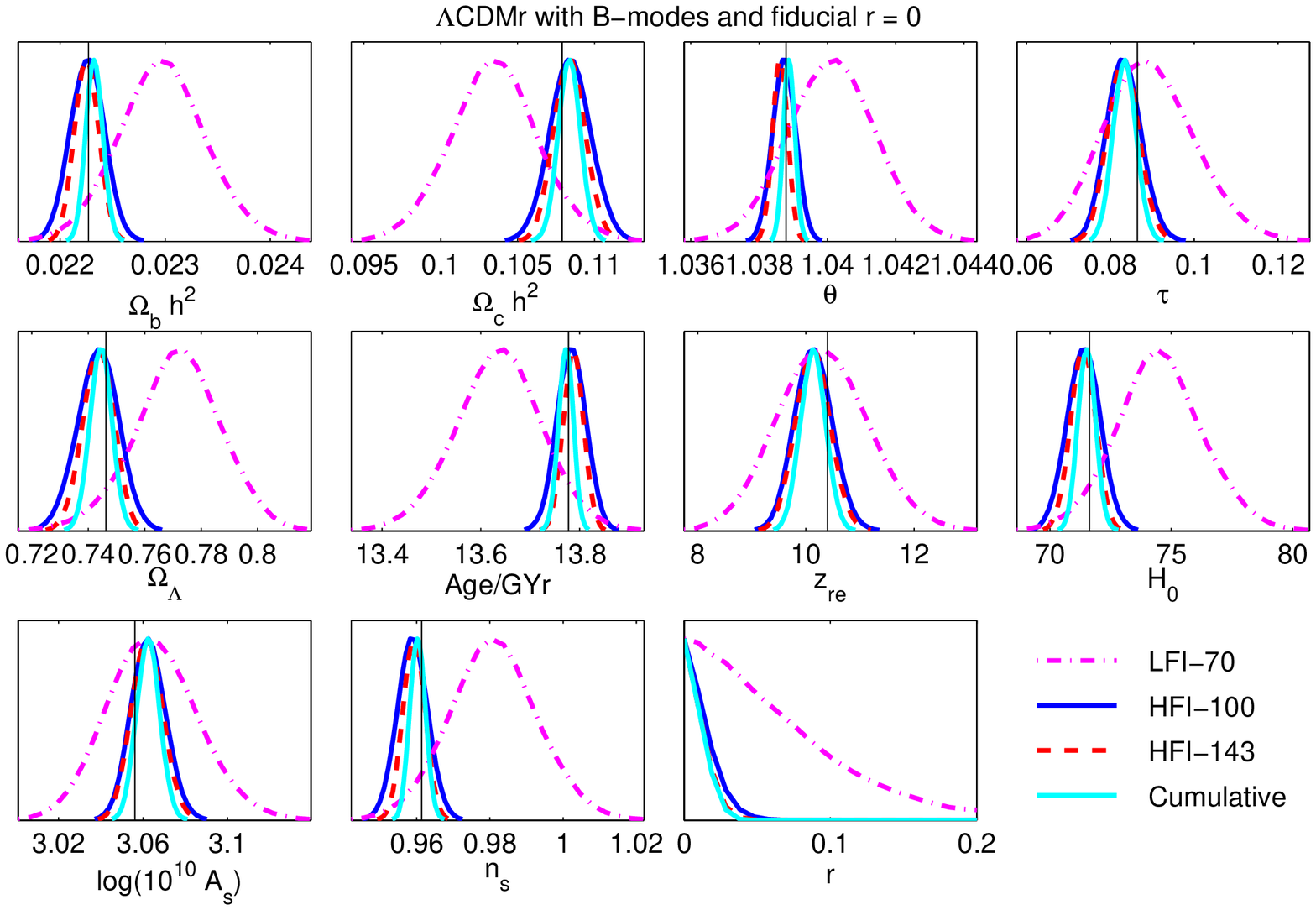} \\
  \includegraphics[height=7cm]{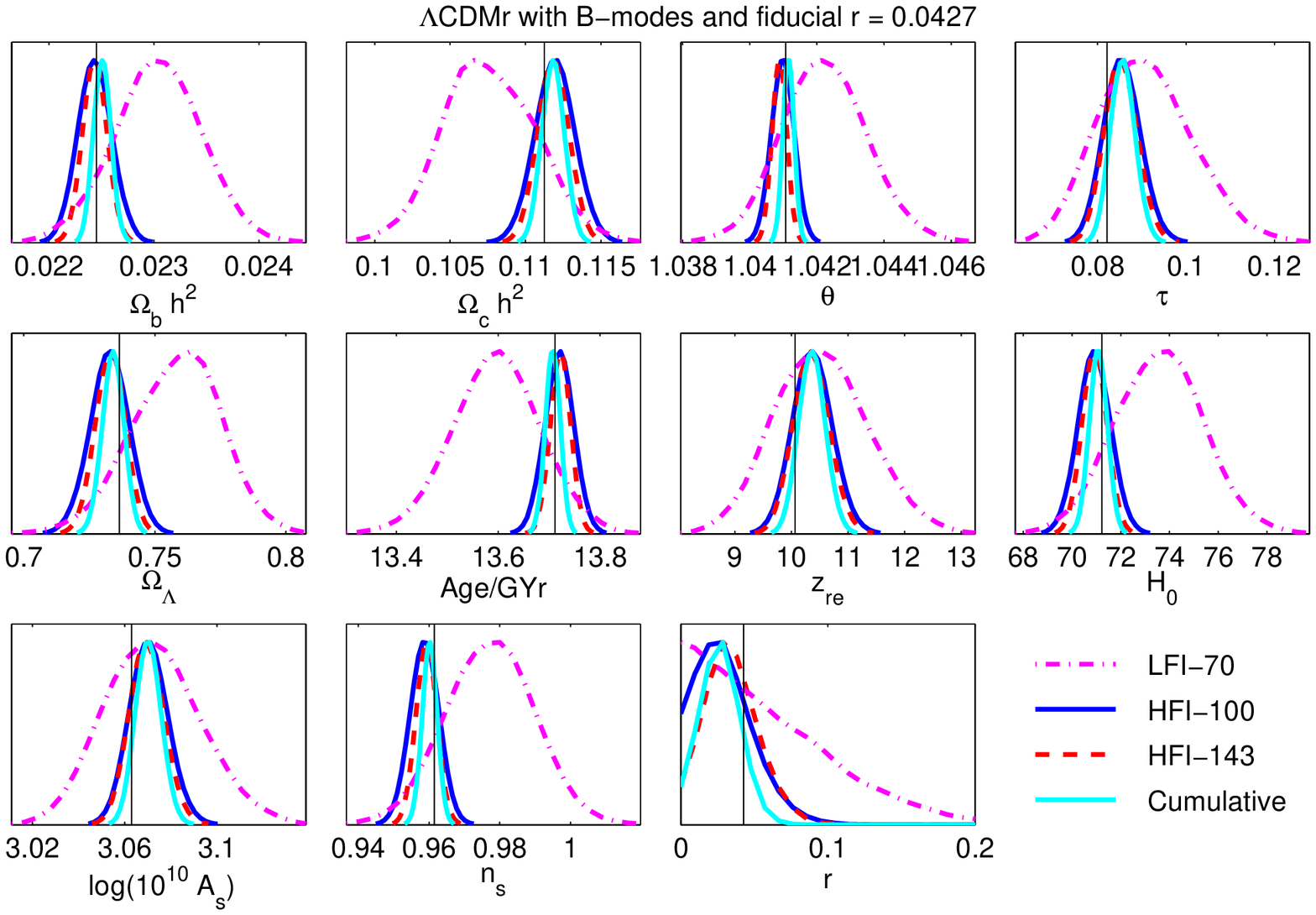}
  \caption{Marginalized likelihood distributions, including $ B $ modes,
    of the cosmological parameters for the $\Lambda$CDM$r$ model. We display the
distributions for each of the three channels and for the cumulative of the three channels.
    The fiducial values are indicated by a vertical thin black line.  The
    fiducial value for the ratio is $ r = 0 $ in the upper panel and $ r = 0.0427 $ in the
    lower panel. Notice that the latter fiducial value is larger than the peaks of the
    marginalized distribution. This is due just to statistical fluctuations
since we are considering only one sky.  
The lower panel figures give upper as well as lower
bounds for $ r: \; \; 0.013 < r < 0.045 $ at $95\%$ CL.
The results on $ n_s $ practically do not change by including or not the $B$
modes as we see from the upper panels of Figs. \ref{r0r04} and \ref{br0br04}.}
  \label{br0br04}
\end{figure}

\begin{figure}
  \includegraphics[height=7cm]{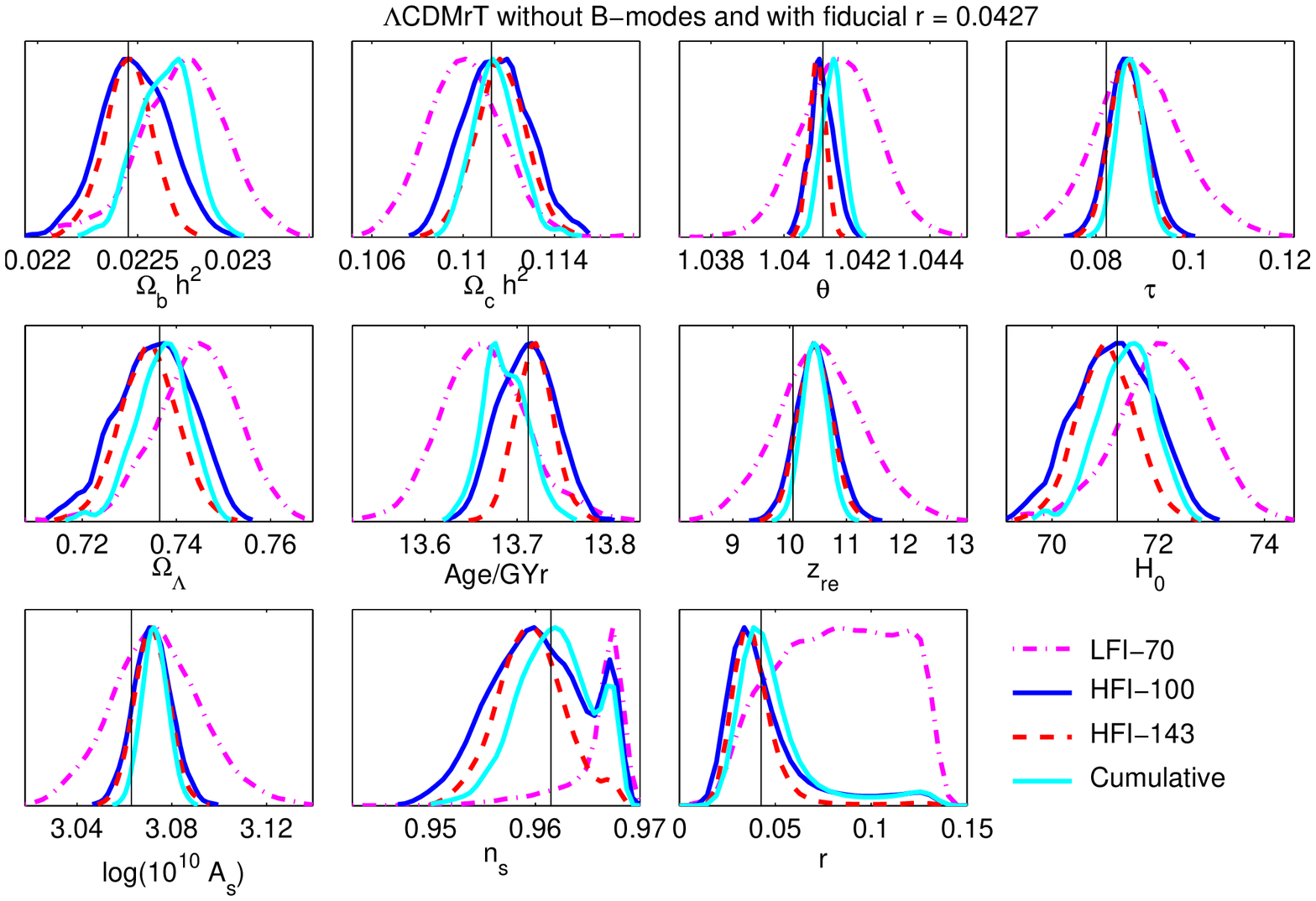} \\
  \includegraphics[height=7cm]{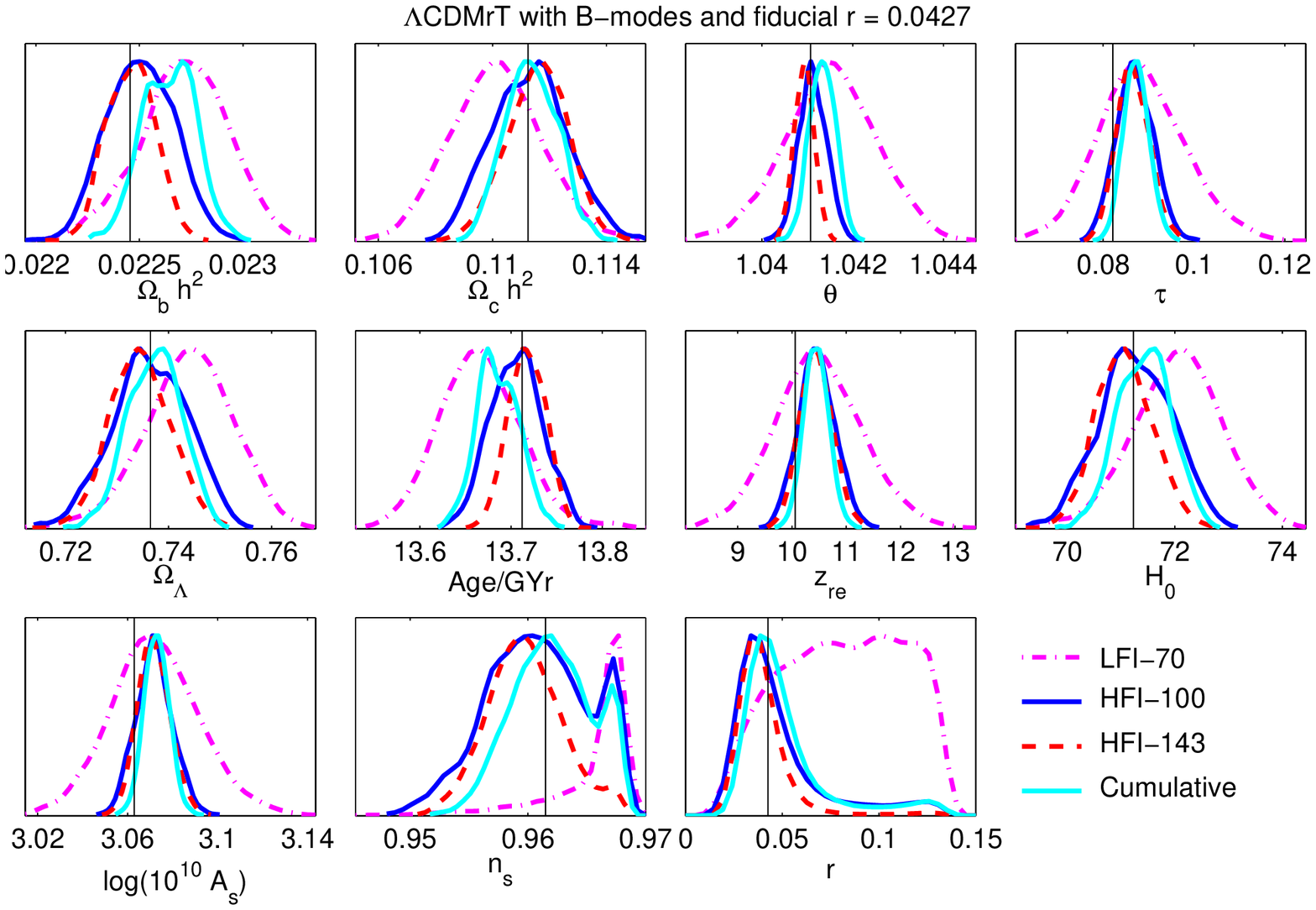}
  \caption{Marginalized likelihood distributions
    of the cosmological parameters for the $\Lambda$CDM$r$T model in which the 
    double--well inflation theoretical model is imposed. 
    The MCMC analysis includes (does not include) $ B $ modes in the lower (upper)
 panel. We display the distributions for each of the three channels and for the 
    cumulative of the three channels. 
The fiducial values are indicated by a vertical thin black line. 
    The fiducial value of $ r $ is well reproduced by the peak of the
    $\Lambda$CDM$r$T distribution in the case of the HFI-100, HFI-143 and
    cumulative  of the three channels both with and without $ B$ modes.
Considerable gain is obtained with respect to the $\Lambda$CDM$r$ model.
Upper and lower panels show quite similar results showing the stability
of the $\Lambda$CDM$r$T model with respect to the inclusion of the $ B $ modes.
Comparison with the lower panel of Fig. \ref{br0br04} shows that
 considerable gain is obtained with respect to the $\Lambda$CDM$r$ model.}
  \label{r04br04T}
\end{figure}

Our findings without including the systematic effects are summarized in 
Figs. \ref{r0r04}-\ref{ban_y} where the marginalized likelihood
  distributions of the cosmological parameters are plotted for several different
  setups. In Tables \ref{tab3} to \ref{tab5}  we list the corresponding relevant numerical values. 
Clearly, in the case of the ratio $ r $, due to the specific form of its
  likelihood distribution, it is more interesting to exhibit upper and lower
  bounds rather than mean values and standard deviations as in Tables \ref{tab3} and  \ref{tab4}.  
  We report the upper bounds and, when present, the lower bounds in 
  Tables \ref{tab5} and \ref{tab6}.  Our conclusions without including the systematic effects are:

\begin{itemize}
\item{The upper bound on $ r $ and the best value of $ n_s $ do not require 
to include the $ B $ modes in the likelihood, and can be obtained with the 
$\Lambda$CDM$r$ model alone, (i.e. $ r < 0.068 $ and $ n_s = 0.9549 $ at 95 \%  CL ).
See Tables \ref{tab3}, \ref{tab5} and Fig. \ref{r0r04}. The inclusion 
of $ B $ modes, for a non vanishing fiducial value, ($ r = 0.0427 $),  allows 
peaked marginalized distributions for $ r $ and a lower bound for $ r $. 
See Table \ref{tab6} and Figs. \ref{r0r04}, \ref{br0br04} and \ref{r04br04T}. 
We obtain $  0.013 < r < 0.045 $ at 95 \%  CL in the 
$\Lambda$CDM$r$ model, with the best values $r = 0.0240, n_s = 0.9597 $.
This shows a substantial progress in the forecasted bounds for $ r $ with 
respect to the WMAP+LSS  data set for which 
$ r < 0.20 $ in the pure $\Lambda$CDM$r$ model \citep{WMAP5,WMAP7}.}
\item{Lower bounds on $ r $ and most probable $ r $ values are always obtained
(with or without the $ B $ modes) with the
$\Lambda$CDM$r$T model. See Tables 
\ref{tab3},  \ref{tab4},  \ref{tab6} and Fig. \ref{r04br04T}. 
The $\Lambda$CDM$r$T model provide  
well peaked distributions for $ r $ on nonzero values $ r \simeq 0.04 $. 
We obtain $ r > 0.039 $ at 68 \%  CL and  $ r > 0.030 $ at 95 \%  CL in the 
$\Lambda$CDM$r$T model.}
\end{itemize}

In Sect. \ref{fortoy} we include in the forecasts the systematic effects discussed in 
Sects. \ref{sensit} and \ref{toym}.
Our conclusions including the systematic effects and foreground residuals are:
\begin{itemize}
\item{The likelihood distributions with and without $ B $ modes result almost the 
same when including the residuals. Only the cosmological parameters 
sensitive to the $B$ modes appear to be affected by the residuals, namely, 
$ \tau, \; z_{re} $ and $ r $.
The main numbers are displayed in Tables  \ref{tab5} and \ref{tab6}.}
\item{The marginalized likelihood $ r $ distribution for fiducial ratio $ r = 0.0427 $ is much
clearly peaked on a value of  $ r $ near the fiducial one in the 
$\Lambda$CDM$r$T model than in the $\Lambda$CDM$r$ model 
(compare Figs. \ref{resbr0br04} and \ref{resbr04T}).
In any case, the best value for $ r $ in the presence of residuals
is about $ r \simeq 0.04 $ (near the fiducial value) both for the 
$\Lambda$CDM$r$ and the $\Lambda$CDM$r$T models.
The $\Lambda$CDM$r$T model turns to be robust, it is very stable 
(its distributions do not change) with 
respect to the inclusion of residuals (and they do not change
neither with respect to the inclusion of $B$ modes). The main numbers are 
included in Tables  \ref{tab5} and \ref{tab6}.
With the $\Lambda$CDM$r$T model
we have for $ r $  at 95 \%  CL:
$$ 
0.028 < r < 0.116 \quad {\rm with ~the ~best ~values} \quad r = 0.04 \quad n_s = 0.9608 \; .
$$}
\end{itemize}

\medskip

It must be stressed that, in the $\Lambda$CDM$r$T model, future improvements in the precision 
$ \delta $ on the measured value of $ n_s $ alone will immediately give an improvement  
$ dr/dn_s \; \delta $ on the prediction for $ r $ as well as for its 
lower bound. Better measurements for $ n_s $ will thus  improve the 
prediction on $ r $ from the $T$, $TE$ and $E$ modes even if a secure detection of
$B$ modes will be still lacking.

\medskip

In order to assess the probability for {\it Planck} to detect $ r $ we also
compute the $B$ mode detection probability by the most sensitive HFI-143 channel; this is done in 
Sect. \ref{143}. We extract $10^5$ skies obtaining the corresponding multipoles $ A_{lm} $
from the $\Lambda$CDM$r$T model according to the procedure described 
in Sect. \ref{mock}, adopting $ r=0.0427 $ as fiducial value.
We compute all the corresponding likelihood profiles only for $ r $ 
and their interesting properties, like
the most likely value $ r_{max} $, the mean value $ r_{mean} $, the standard 
deviation $ \Delta r_{\max} $ of the $ r_{max} $ distributions, 
the skewness and the kurtosis,
(which measures the departure from a Gaussian  likelihood), Fig. \ref{probd}.
We finally compute the 99\% CL, 95\% CL, and 68\% CL lower bounds
for $ r $.  The probabilities of detection of $ r $ are displayed in Fig. \ref{probd2}. 
At the level of foreground residual equal to 30\% of the considered toy model, 
only a 68\% CL (one sigma)
detection is very likely. For a 95\% CL detection (two sigmas)
the level of foreground residual should be reduced to 10\% of the 
considered toy model, or lower.

Lensing acts on the B-modes as a contamination by transforming E-modes
into B-modes. It is a frequency independent effect while residuals are 
frequency dependent. Lensing weakens the signal around $ \ell \sim 90 $ 
where the primordial B-modes peak but not in the small $ \ell $ modes range where 
the reionization bump dominates. On the other hand foreground 
residuals are larger at small $ \ell $ than at $ \ell \sim 90 $. Namely, 
residuals and lensing affect the detection of B-modes
in complementary ways, with the effect of residuals stronger than 
that of lensing. As a consequence, lensing plus residuals
can spoil the detection of $ r $ even when residuals are assumed
at the 30\% level of the considered toy model. 
On the contrary, lensing in the absence
of residuals still allows a detection of $ r $. For example, 
several MCMC simulations show that our lower
bounds on $ r $ are not significantly affected by lensing in the absence 
of foreground residuals. Let us make clear, at any rate, that lensing 
was not considered in the analysis of the $r-$detection probability in Sect. 
\ref{143}. Finally, it should be clear that if the theoretical 
constraint $ r = r(n_s) $ of the $\Lambda$CDM$r$T model is imposed 
on the MCMC analysis, $ r $ has always well defined lower bounds 
regardless of lensing and/or residuals.

The forecasted probability of detecting $ r $ 
is based on the statistics of the
shape of the $r$ -likelihood. This shape determines whether a detection of
$ r $ can be claimed with a given confidence level.
But real CMB experiments can observe only 
one sample: the observed sky. So, the possibility of inferring $ r $ 
from one single (albeit very large) sample depends on the 
sample itself, and therefore, whether $ r $ will be or will be not 
detected depends also of a question on luck.

In addition, the results for many skies presented 
in Sect. \ref{143} show the consistency of our whole approach to determine $ r $.
%Similar results are valid for other cosmological parameters.

\medskip

Finally, in Sect. \ref{efbias} we consider the bias 
effect in the foreground residuals implemented as a linear
perturbation affecting the $ C_l'$s and explore how the cosmological parameter
distributions are affected by the bias. We implement two extreme cases: in case (i)
the bias fluctuates randomly around zero and in case (ii) the bias fluctuates around a
non-zero value, staying significantly non-zero. In case (i) the cosmological parameters
are practically unaffected while in case (ii)  the peaks of the cosmological parameter
distributions are shifted within one or two sigmas of the WMAP values. 
In particular, $ r $ is not anymore detected in case (ii).

\medskip

The best and mean values reported here for $ r $ and the
other cosmological parameters do not correspond to the true sky data
but to mock skies generated from the MCMC simulations as explained above.
Nevertheless, the deviations between the best and the fiducial values are
relevant indicators for $ r $ as well as the lower and upper bounds and the standard deviation.
The fact that the fiducial and mean values of  $ r $ are very close and that $ \Delta r_{\max} $
coincides with the mean value of the standard deviation of $ r $
indicate that {\it Planck} can provide detections of high quality.

More in general, our results support the quest  for $B$ mode polarization  
in the current CMB data and future $B$ oriented polarization missions
under study by both ESA\footnote{http://www.b-pol.org/index.php} and NASA\footnote{http://cmbpol.uchicago.edu/}
\citep{2009ExA....23....5D,2006AAS...209.4907B}.

\section{Fitting current CMB + LSS data with the Ginzburg-Landau 
effective theory of Inflation}
 \label{LGtheory}
 
As discussed in the introduction, 
the effective theory of inflation within the G-L approach gives precise
order of magnitude estimates for the spectral index $ n_s $, the ratio of tensor to scalar 
fluctuations $ r $ and the running of the spectral index $ dn_s/d \ln k $ \citep{reviu}.

\medskip

\begin{figure}
\includegraphics[height=10cm]{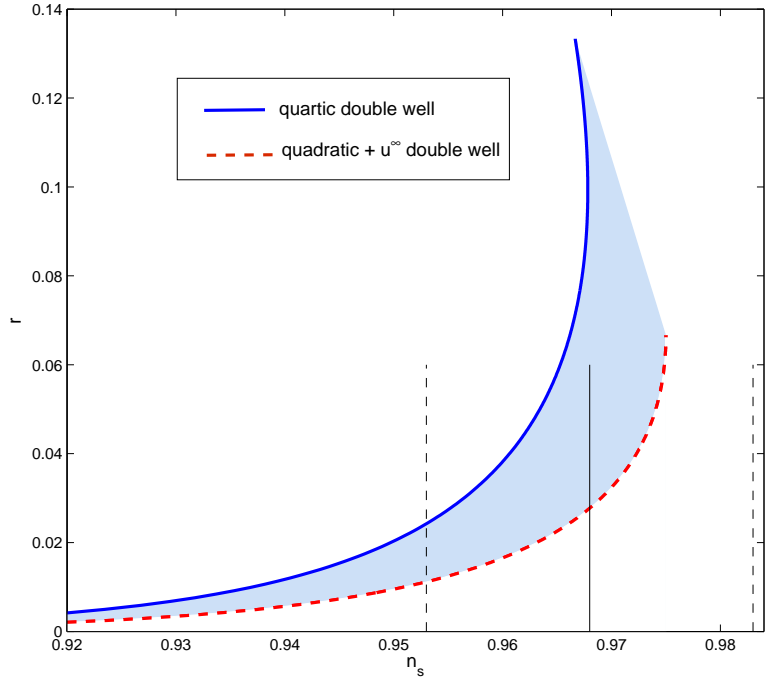}
\caption{The universal banana region $ \cal B $ in the $ (n_s, r) $-plane
  setting $ N = 60 $. The upper border of the region $ \cal B $ corresponds to
  the fourth order double--well potential expressed by Eq. (\ref{binon}).  
The lower border is
  described by the potential $ V(\varphi) = \frac12{ m^2} \,
  \left(\frac{m^2}{\lambda} - \varphi^2\right) $ for $ \varphi^2 < m^2/\lambda $
  and $ V(\varphi) = \infty $ for $ \varphi^2 > m^2/\lambda $ \citep{high}.  We
  display in the vertical full line the $\Lambda$CDM$r$ value $ n_s = 0.968 \pm
  0.015 $ using the WMAP+BAO+SN data set.  The broken vertical lines delimit
  the $ \pm 1 \, \sigma$ region.}
\label{banana}
\end{figure}

Within the context of the G-L effective theory of inflation, the work in 
\citet{1sN}, \citet{mcmc1}, \citet{mcmc2}, \citet{high}, \citet{reviu}  
showed that:
\begin{itemize}
\item The small inflaton selfcoupling arises naturally as the ratio of the
  inflation energy scale and the Planck energy. The inflaton mass is small
 compared with the inflation energy scale.
\item The amplitude of the CMB anisotropies sets the energy scale of inflation
  to be $ M \sim 10^{16} $ GeV for all generic slow-roll
  inflationary potentials.
\item Double-well inflaton potentials give the best fit to CMB+LSS data.
  Basically, the inflaton potential must have a negative second derivative at
  horizon exit which favours double-well potentials over single well potentials.
\item For double-well quartic inflaton potentials, the best value for the ratio tensor
  to scalar fluctuations is $ r \simeq 0.05 $ with the lower bound 
$ r > 0.023 \;(95\%)$ CL in the case of the quartic double-well potential.
The novelty in the MCMC analysis of the CMB+LSS data that leads to these results
is in the fact that we imposed the analytical expressions for $
  n_s $ and $ r $ derived from the inflaton potential as a hard
    constraint \citep{mcmc1}.
\item Higher order double-well inflaton potentials are investigated in
  \citet{high}.  All $ r=r(n_s) $ curves for double--well even potentials of high
  order fall inside a universal ``banana-shaped'' region $ \cal B $, Fig. \ref{banana}.
\end{itemize}

The fourth order binomial potential provides the simplest double--well potential
best reproducing the CMB+LSS data within the G-L effective theory approach: 

\begin{equation}\label{binon}
  V(\varphi) = \frac{\lambda}{4} \left( \varphi^2- 
    \frac{ \; m^2}{\lambda} \right)^2 ~~~~, ~~~~~~\lambda = \frac{y}{8 \, N}\left(
    \frac{M}{M_{PL}}\right)^4, ~~~~~~m \equiv \frac{M^2}{M_{PL}}
\end{equation}
where $ \varphi $ is the inflaton field$, \lambda $ stands for the quartic
coupling, ($ y $ being the corresponding coupling of
  order one), $ m $ is the inflaton mass. Notice that the quartic coupling $
\lambda $ is proportional to the ratio $ M/M_{PL} $ to power four and hence very
small as stated above for all the inflaton self-couplings.

Adding higher order terms with additional parameters does not improve significantly
the fits \citep{high}.

\medskip

In \citet{high} it is found that the $ r=r(n_s) $ curves for double--well inflaton 
potentials in the G-L spirit fall inside the universal
banana region $ \cal B $ depicted in Fig. \ref{banana}.

\medskip

The lower border of the universal region $ \cal B $ is particularly relevant
since it gives a lower bound for $ r $ for each observationally
allowed value of $ n_s $. For example, the best value $ n_s = 0.964 $
implies from Fig. \ref{banana} that $ r > 0.021 $.
The upper border of the universal region $ \cal B $ tells us the upper bound
$ r < 0.053 $ for $ n_s = 0.964 $.
Therefore, we have within the large class of potentials inside the region $ \cal B $
$$
 0.021 < r < 0.053 \quad {\rm for}\quad  n_s = 0.964 \; .
$$
Moreover, the fourth order double--well potential represented by Eq. (\ref{binon}) is the simplest
and G-L stable inflaton potential reproducing very well 
the present CMB+LSS data. 

\medskip

Not using the Ginsburg-Landau approach, the lower bounds for $ r $
are not obtained. \cite{kinney08,PeirisEasther}  do not use the Ginsburg-Landau 
approach, do not find lower bounds for r and cannot exclude values for $ r $ much smaller 
than $ r = 0.0427 $. 

It must be noticed that our present analysis shows that values $ r \ll 0.0427 $,
(and hence very small B modes) are outside the possibilities of detection by Planck.

\medskip

Future improvements on the precise value of $ n_s $ alone will immediately give an improvement
on the theoretical prediction for $ r $ as well as for its lower bound.
An improvement $ \delta $ on the precision of $ n_s $ implies
an improvement 
$$ 
\frac{dr}{dn_s} \; \delta 
$$ 
on the precision of $ r $. 
According to \citet{high}, at $ n_s = 0.964 $ from $ r=r(n_s) $ we have:
\noindent 
$$ 
\frac{dr}{dn_s} = 4.9 \; \; {\rm on ~ the ~ upper ~ border ~ of} \; 
{\cal B}  \;  ({\rm fourth ~ degree ~ double ~ well}) \; \; 
 {\rm and } \;  \; $$
\noindent 
$$\frac{dr}{dn_s} = 1.35 \; \; {\rm  on ~ the ~ lower ~
border ~ of} \;   {\cal B} .
$$
Better values for $ n_s $ will thus improve the prediction on $ r $ from the $T$, 
$TE$ and $E$ modes while a secure detection of $B$ modes is still lacking.

\section{{\it Planck} sensitivity}\label{sensit}

In this study we exploit the {\it Planck} sensitivity and resolution at its three favorite cosmological 
channels, i.e. at the frequencies of 70, 100, and 143 GHz.

Table \ref{table:sens} reports the {\it Planck} performance at these frequencies, based on 
\citet{PlanckBlueBook} but consistent with the most recent pre-launch measurements 
of the HFI channels at 100 GHz and 143 GHz \citep{lamarre2010,Maffei_prelaunch}, 
and, for the LFI channel at 70 GHz, as updated in \citet{prelauM}, \citet{prelauB}, \citet{sandri_etal_2010}.
%HFI performance at 100 and 143 GHz is in agreement with \citep{PlanckBlueBook}.
Notice that the LFI sensitivity reported here includes also the fluctuations of the 4K reference 
load, since it is obtained on ground-based calibration performed under realistic conditions. 
The resolution at the various frequencies comes from accurate optical simulations.
Also notice that these numbers are likely conservative, i.e. in principle a further refinement of 
tuning could return into an improvement of in-flight sensitivity.
Almost four surveys have been adopted in this work.

\section{Residuals from systematic effects and foregrounds: toy model}
\label{toym}

The sensitivities presented above do not include the degradation in accuracy that could come from 
various sources of systematic effects, of both instrumental and/or astrophysical origin.

In this section, we discuss the current estimates publicly available for the systematic effects 
affecting {\it Planck} measurements.
 
\subsection{Straylight}

{\it Planck} achieves very good side lobe rejection thanks to its telescope design 
\citep{2004A&A...428..299S,sandri_etal_2010,Maffei_prelaunch,tauber_etal_2010}. 
In spite of this, the main source of contamination at large angular scales or at low 
multipoles comes from the so-called 
straylight effect, i.e. the signal entering in the lobes at various angular 
distances from the main beam. 
It can be distinguished in straylight from the intermediate beam, i.e. at angular 
distance of few degrees 
from the main beam, and from the far beam, i.e. at angular distance from the main 
beam larger than 
some degrees. The main sources of straylight are the Galactic emission and the CMB dipole.    

The straylight from the intermediate beam introduces a sort of smearing of signal around 
that observed by 
the main beam. Detailed studies show that it is important only 
close to the Galactic plane, 
a region typically 
excluded from scientific analysis through suitable masks, while it is significantly 
less of the straylight 
from the far sidelobes in all the other sky regions, and, ultimately, for the 
recovery of the CMB angular 
power spectrum. We will then consider only far sidelobes in the following 
estimates.
 
Notice that, if the optical behaviour is well known it is possible to subtract to high precision 
 this effect from the data by simply evaluating it on the observed sky by means of convolution codes 
taking into account the effective observational strategy, with the only (small) limitation introduced 
by the receiver noise. In practice, this correction is limited by the accuracy in the knowledge of 
optical behaviour. We will assume for numerical estimates an effective uncertainty of 
$ \sim 30 $\%  in the beam response in the sidelobes, implying that the amplitude of the spurious 
effect remaining in the data is about one order of magnitude smaller than the original effect. 

Therefore, assuming the (conservative) sidelobe levels computed for {\it Planck} at 
frequencies of 70 and 100 GHz,  
a reasonable estimate for the residual straylight from the Galaxy, 
rescaled from the computations carried 
out for the original effect \citep{2001A&A...373..345B, 2004A&A...428..311B,sandri_etal_2010}, is 
\begin{equation}
C_\ell \;  \;   \frac{\ell \; (\ell+1)}{2 \, \pi} \sim 8 \times 10^{-4} \; \mu K^2 
\quad {\rm for} \quad \ell \le 10 \; ,
\end{equation}
\noindent
and
\begin{equation}
C_\ell  \;  \;  \frac{\ell \; (\ell+1)}{2 \, \pi} \sim 2.5 \times 10^{-4} \;  \mu K^2 
\quad {\rm for} \quad \ell \ge 11 \; .
\end{equation}

Notice that at frequencies $ \nu \ge 70 $ GHz dust emission is the main Galactic
diffuse foreground, while at 30 and 44 GHz also free-free and synchrotron emission are relevant.
Of course, the straylight effect is larger at 30 and 44 GHz but we could neglect
them in this study for cosmological parameter estimation.
Assuming similar sidelobe levels at 143 GHz, given the typical dust frequency 
behaviour at millimeter wavelengths $ T_\nu $ almost proportional to $ \nu^2 $, we have:
\begin{equation}
C_\ell  \;  \;  \frac{\ell \; (\ell+1)}{2 \, \pi} \sim 3.2 \times 10^{-3} \;  \mu K^2 
\end{equation}
\noindent
for $ \ell \le 10 $, and
\begin{equation}
C_\ell  \;  \;  \frac{\ell \; (\ell+1)}{2 \, \pi} \sim 1 \times 10^{-3}  \; \mu K^2 
\end{equation}
\noindent
for $ \ell \ge 11 $.

\medskip

The other relevant source of straylight contamination is the CMB dipole 
\citep{2006MNRAS.371.1570B,2007MNRAS.376..907G}. Note that, for symmetry reasons, 
this effect is significant only at even multipoles while it is negligible 
in practice at odd multipoles. 
Again, rescaling the results obtained for the original effect, we derive a 
suitable range for the estimate 
of the residual contamination   
\begin{equation}
C_\ell \;  \; \frac{\ell \; (\ell+1)}{2 \, \pi} \sim 0.016 \div 0.16 \;  \mu K^2 
\end{equation}
for even multipoles (with a typical value of $ 0.048  \, \mu K^2 $) and 
\begin{equation}
C_\ell \;  \; \frac{\ell \; (\ell+1)}{2 \, \pi} \simeq 0 
\end{equation}
for odd multipoles. The larger values apply at lower multipoles (up to about 10),
the lower  ones at higher multipoles. 
Although the exact value depends on the particular considered receiver, 
we assume these estimates constant with frequency, 
as in the case of sidelobe levels approximately constant with frequency,  
being the CMB signal frequency independent  (in equivalent thermodynamic temperature).

Notice that, at the frequencies of 70$\div$143 GHz, 
for even multipoles dipole straylight is larger than Galactic straylight.
 
In general, the contamination from straylight can be modeled to first approximation
as an additional spurious excess of power. 
In principle, one could also include perturbations multipole by multipole (or multipole band by 
multipole band) of the above estimates to avoid a modeling in terms of a simple analytical form
for the spurious additional power.

\medskip

The straylight effect in polarization mainly depends on the (non-perfect) balance of the straylight 
in total intensity in the coupled receivers used to extract the Q and U Stokes parameters 
\citep{2004A&A...428..311B}.
On the basis of optical simulations, we could assume relative differences of few tens per cent, 
which should reduce of about one order of magnitude the original effect with respect to that 
in total intensity. On the other hand, the modeling and verification of optics in polarization 
is much more complex than in total intensity. We then expect that it will be more difficult to 
use optical predictions to subtract this effect into the data and assume a residual effect similar 
to the original one. Therefore, we estimate that the amplitude of the residual effect in polarization 
will be similar to that  in total intensity.

\subsection{Main beam asymmetry, leakage, time constants, and glitches}

Another potential systematic effect that needs to be kept under control is the 
effect of the antenna beam profile in the main beam. Beam profiles
exhibit a deviation from perfect circular symmetry, 
in the range from a few percent up to $\sim 30$\% in the case of the lowest frequency channels 
\citep{sandri_etal_2010,Maffei_prelaunch}. Main beam distortions are in principle 
a source of concern as they can bias the estimated power spectra in the high 
$\ell$ regime, and hence affect the likelihood models and cosmological parameters. 
This happens for two reasons. {\it Planck}'s scanning strategy is not isotropic but has 
a preferred direction, roughly coincident with ecliptic meridians. 
In the first place, this fact makes the $ \ell $ space equivalent window function 
of the beam rather non trivial, and difficult to  estimate analytically (though 
approximate analytical solutions do exist, see e.g. \citet{2002PhRvD..65f3003F}). 
Secondly, and more important, 
{\it Planck} estimates the Stokes linear polarization parameters by combining measurements 
taken from different detectors. The beam asymmetry renders the contribution to the 
intensity $ I $ unbalanced even when the same pixel is observed, because of the 
different orientation and shape of the beams. In turn, this can create $ I $ to 
$ Q, \; U $ leakage \citep{2009A&A...493..753A} and produce biases in the polarization power spectra.

Fortunately, the beam profiles for {\it Planck} have been measured very well 
during ground testing campaigns and will be cross 
checked in flight (see \citet{2000astro.ph.12273B}, \citet{2007AstBu..62..285N}, 
\citet{2010A&A...510A..58H} and references therein). 
Furthermore, analytical and semi-analytical machinery exist 
to estimate the $ \ell $ equivalent window function asymmetric beams, 
once a beam profile is known and a scanning strategy assumed. These methods 
compute, for each multipole $ \ell $ the beam coupling matrix between all power spectra 
(thus taking leakage into account), starting from an approximate model of the 
scanning strategy \citep{2009A&A...493..753A} that can be refined performing 
signal-only Monte Carlo simulations for the CMB component \citep{varenna}. 
While a thorough analysis of the accuracy of these procedures have not been 
performed yet, it is fair to expect that main beam distortions will not be a 
major source of systematic contamination for {\it Planck} (see e.g. \citet{2009arXiv0907.5254R}). 

Not even a satellite experiment as Planck can safely assume to use the 
entire sky for CMB analysis. Incomplete sky coverage can induce leakage 
of the E polarization modes into B modes if a sub-optimal power spectrum 
estimator is employed. While this effect is not connected to the beam,
but rather of geometrical origin, it is worth mentioning here because 
it may trigger spurious detection of B modes. In a realistic analysis, 
the leakage effect is corrected from the beginning by using pseudo 
$ C_\ell $ methods \citep{master} that are the standard choice for power 
spectrum estimation in the high $ \ell $ ($ \gtrsim 30 $). Pseudo $ C_\ell $ 
methods correct for leakage by means of coupling kernels: in particular, 
the E and B mode pseudo-spectra exhibit correlations that need to be  
accounted for (see, e.g., appendix A in \citet{kogut}). At low multipoles,
pixel based methods are normally used to compute directly the likelihood 
function without assuming power spectrum estimation as an intermediate 
step. Pixel based methods do not suffer from leakage \citep{tegmark_qml}.

The bolometric detectors of HFI exhibit a non trivial transfer function, that
distorts the signal both in amplitude and in phase.  Qualitatively, the effect on
amplitude is akin to a first order low pass filter arising from the detector's
intrinsic time constant modified by electro-thermal effects \citep{lamarre2010}.
Knowledge of the filter function allows one to deconvolve out its effect on
timelines at the price of increasing (slightly) and distorting the high frequency noise
level, which becomes non white (`colored noise'). These measurements will be
performed in flight \citep{lamarre2010}. Any residual error
would have an impact similar to beam smearing along the scan direction, so it
contributes to beam asymmetry which, as stated above, can be accounted for with high confidence. 

Another potential source of concern for bolometric detectors is that they are sensible to
cosmic ray hits, that create glitches in the timeline, i.e.  they are always seen
as positive spikes in the bolometer signal \citep{lamarre2010}, followed by a tail
also due to the bolometer's time constant. These events can be detected and flagged
on the timelines. Residual effects (due to undetected glitches) can be kept under
control \citep{2010arXiv1001.2509M} for the cosmological analysis by relying on
angular power spectra obtained cross correlating different detectors (also known as
cross-spectra see e.g. \citet{2005JCAP...11..001P}).

\subsection{Residuals from foregrounds}

The most important source of contamination in a CMB experiment like {\it Planck} 
will come from the residual of astrophysical foregrounds. In fact, although the wide {\it Planck} 
frequency coverage, possibly complemented by WMAP maps and ground-based 
and balloon-borne experiment data, is 
particularly advantageous for a precise removal of astrophysical signals from the maps
and the accurate mapping of CMB anisotropies, nevertheless we expect 
that a certain level of residual 
contamination will remain into CMB maps, particularly in polarization. 
Many methods of component separation, each with its own pros and cons, 
have been and are continuously 
elaborated for the analysis of {\it Planck} multifrequency maps 
(see \citet{2008A&A...491..597L} and references therein). 
In the present work, we are interested in 
the residuals from astrophysical foregrounds affecting the recovery 
of the CMB angular power spectrum. 
It is typically given as a difference between the input CMB angular 
power spectrum and the CMB angular 
power spectrum estimated after the component separation layer.
In general, it is not so meaningful to provide a description of 
foreground residuals different at different frequencies, since, 
by definition, the component separation 
layer exploits exactly the multifrequency mapping of the sky. 
Thus, the estimate adopted 
in this work have to apply to the whole set of frequency channels.

\medskip

Different methods show different residuals at various ranges of multipoles. 
The multipole dependence, or, in other words, the shape of this residual also depends 
on the considered method.   

Concerning residuals for the $T$ mode, recent simulations \citep{2008A&A...491..597L}  show 
residual shapes only slightly dependent on the multipole, with amplitudes in the range
\begin{equation}
C_\ell  \;  \frac{\ell (\ell+1)}{2 \, \pi} \sim \frac{10^2 \div 10^3}{2 \, \pi} \,  \,  \mu K^2 \, ,
\end{equation}
\noindent 
the exact value depending on the method and, for each method, on the particular multiple
band, with typical variations of about 30-40\%. In this work, we model this $T$ spurious power
as flat in $ C_\ell  \; \ell  \; (\ell+1) $.

\medskip

Galactic polarized foreground (mainly from diffuse synchrotron and dust emission) affects
CMB angular power spectrum recovery more significantly in polarization than in temperature. 
We expect that their residual after component separation 
will take partially memory of the original shape of the 
foreground power spectrum, in particular at large scales where they show much more power than the CMB.
Again, different methods give different residuals, 
regarding both multipole dependence and amplitude.
In this work, we model the foreground residual for the $TE$ mode and $E$ and $B$ modes 
(assumed to be equal, $B$ = $E$) 
as the sum of two shapes, the first one  (dominant at low multipoles)  
described by the foreground shape properly rescaled in amplitude, the second one constructed from 
the foreground shape properly rescaled in amplitude and changed in slope. 

Fig. \ref{foreres_TE_EE} displays our ``starting'' 
conservative models for the residuals in $TE$ mode and in 
polarization modes. 
We let us the freedom to simply rescale them with multiplicative factors in order to address 
typical level of foreground residuals for which the impact on our cosmological aim is not critical 
(see also Sect. \ref{subsec:param}).

\subsection{Additional noise versus bias}

All the systematic effects discussed above, coming from instrumental effect, sky signal, or from their 
coupling, can be considered in two different schemes. 

In the first, simplest case, they can be treated as sources 
of spurious additional noise power, i.e. they 
do not introduce a bias affecting the recovery of the estimation of the 
CMB angular power spectrum but they increase our uncertainty in its recovery. 
Therefore, the effect can be modeled adding in quadrature the 
quoted $ C_\ell $ of the power of the residual systematics 
to those coming from sensitivity, resolution, and 
cosmic plus sampling variance. This approach is equivalent to 
assume that we will be able to properly model 
and subtract a correct estimation for the systematic effects so that only a statistical uncertainty in 
their subtraction will affect the data. 

In another, more critical approach, one can assume to miss the correct 
estimation of the spurious effects.  
Their systematic effects will be then much more dramatic, i.e. they will introduce 
also a bias in the estimation of the CMB angular power spectrum.
This case can be modeled ``perturbing'' the $ C_\ell $ to be compared with the exact model
linearly adding the additional spurious power as described above.

\subsection{Parametric approach to systematic effects}
\label{subsec:param}
 
In general, we do not use in this  work a precise (still not completely available) description
of the considered systematic effects, but only suitable representations of them.
Therefore, we will use our estimations to understand if the considered classes of systematic effects
may significantly affect the cosmological exploitation of {\it Planck} data with respect to the 
determination of cosmological parameters possibly by rescaling the estimation quoted above.
This is done with the aim of identifying the corresponding levels at 
which it is necessary to control the systematic effects in order to avoid to spoil 
the scientific accuracy of the {\it Planck} data.  %demanded to fix this problem. 
We technically implement this rescaling with a multiplicative constant on the 
residuals of systematic effects on the $ C_\ell $ described in the previous sections. 

\section{Mock data production and likelihoods}
\label{mock}

We describe in this section the theoretical basis of our simulations when
experimental errors are treated as statistical noise. This includes the
instrumental white noise as well as the residuals from systematic errors and foregrounds as 
described in the previous section. 
In other words, we assume that the noise contribution to the observed
CMB skies due to systematic errors can be precisely assessed, thanks to suitable 
procedures such as cleaning simulations in case of foreground residuals.

\medskip 
Let us denote the fiducial theoretical multipoles as $ {\hat C}_\ell^{TT}, \;
{\hat C}_\ell^{TE}, \; {\hat C}_\ell^{EE}, \;{\hat C}_\ell^{BB} $ and the
(possibly $\ell-$dependent) noise covariances as $ N_\ell^{TT}, \; N_\ell^{EE} $
and $ N_\ell^{BB} $. For instance, in the case of foreground residuals, we would have
\be \label{covres}
  N_\ell^{XX'} = w_\ell \; R_\ell^{XX'} + n^{XX'} \;,\quad X,X' = T,E,B
\ee
where $ w_\ell $ is the window function in multipole space, $ n^{XX'} $ is the
white instrument noise and $ R_\ell^{XX'} $ are appropriate quantities which can
be estimated through map cleaning simulations.  In any case we assume here that
$ N_\ell^{BB}=N_\ell^{EE} $ and that $ N_\ell^{TE}=N_\ell^{TB}=N_\ell^{EB} = 0 $.

\medskip
Thus, the full covariances of the $ T-E $ fluctuations read
\begin{equation}
  \begin{pmatrix}
    w_\ell\;{\hat C}_\ell^{TT} + N_\ell^{TT} &  w_\ell\;{\hat C}_\ell^{TE} \\
    w_\ell\;{\hat C}_\ell^{TE} & w_\ell\;{\hat C}_\ell^{EE} + N_\ell^{EE}  \\
  \end{pmatrix} 
  = {\cal R}_{\ell} 
  \begin{pmatrix} 
    {\hat{\cal C}}_\ell^{+} & 0 \\
    0 & {\hat{\cal C}}_\ell^{-} \\
  \end{pmatrix}
  {\cal R}^{\,\rm t}_{\ell} 
\end{equation}
where $ {\cal R}_{\ell} $ are suitable rotation matrices and 
$ {\cal R}^{\,\rm t}_{\ell} $ stands for the trasposed matrix of $ {\cal R}_{\ell} $.

The $ B $ fluctuations are decoupled, that is $ {\hat C}_\ell^{BE}={\hat
  C}_\ell^{BT} = 0 $ and have full covariance $w_\ell\;{\hat C}_\ell^{BB} +
N_\ell^{BB} $.

\medskip

With these notations, a possible observed set of fluctuation amplitudes reads 
\begin{equation}
  \begin{pmatrix} 
    A_{\ell m}^{T} \\ 
    A_{\ell m}^{E^{\phantom{I}}} \\ 
  \end{pmatrix}
   = {\cal R}_{\ell}
  \begin{pmatrix} 
    \sqrt{{\hat{\cal C}}_\ell^{+}} \; g_{\ell m}^{+} \\ 
    \sqrt{{\hat{\cal C}}_\ell^{-}} \; g_{\ell m}^{-} \\  
  \end{pmatrix}
  \;,\quad A_{\ell m}^{B} = \sqrt{w_\ell\;{\hat C}_\ell^{BB} + N_\ell^{BB}}\;g_{\ell m}^{B}
\end{equation}   
where $ g_{\ell m}^{X}, X \equiv +,-,B $, are independent centered unit Gaussians, 
that is 
\begin{equation}
  \avg{g_{\ell m}^{X}} = 0 \;,\quad
  \avg{g_{\ell m}^{X}\;g_{\ell' m'}^{X'}} = 
\delta_{\ell\ell'} \; \delta_{mm'} \; \delta^{XX'}
\end{equation}
The amplitudes $ A_{\ell m}^{X} $ are assumed real (which is always possible
for integer weights $ \ell $).

\medskip

The corresponding observed multipoles $ \overline{C}_\ell $
(sometimes called pseudo-$C_\ell$) are
\begin{equation}
  \overline{C}_\ell^{XX'} = \frac1{2\ell+1}\sum_m A_{\ell m}^{X}\; A_{\ell m}^{X'}
\end{equation}

Consider now the multipoles $ C_\ell^{XX'} $ produced by some test cosmological
model. In the approximation exploited in this work (all-sky coverage and uniform sensitivity), 
the likelihood of such multipoles, given the observed
$ \overline{C}_\ell^{XX'} $, can be written as
\begin{equation}
  L = \exp\left( { -} \tfrac12 \chi^2_{T,E}  - \tfrac12 \chi^2_B \right)
\end{equation}
where
\begin{equation}\label{chi2}
  \begin{split}
    \chi^2_{T,E} &=  \sum_\ell (2\ell+1) \left[ {\rm tr}\left(
        C_\ell^{-1} \overline{C}_\ell\right) - 
      \log\frac{{\rm det}\overline{C}_\ell}{{\rm det}C_\ell} -2 \right]\\
    \chi^2_B &= \sum_\ell (2\ell+1) \left( x_\ell - \log x_\ell -1\right)
  \end{split}
\end{equation}
and $ C_\ell , \; \overline{C}_\ell $ are the $2\times 2$ matrices 
\begin{equation}\label{Candx}
  C_\ell = 
   \begin{pmatrix}
    w_\ell\;C_\ell^{TT} + N_\ell^{TT} &  w_\ell\;C_\ell^{TE} \\
    w_\ell\;C_\ell^{TE} & w_\ell\;C_\ell^{EE} + N_\ell^{EE}  \\
  \end{pmatrix} 
  \quad,\qquad \overline{C}_\ell = 
  \begin{pmatrix}
    \overline{C}_\ell^{TT} & \overline{C}_\ell^{TE} \\
    \overline{C}_\ell^{TE} & \overline{C}_\ell^{EE^{\phantom{I}}} \\
  \end{pmatrix}
\end{equation}
while 
\begin{equation}
  x_\ell = \frac{{\bar C}_\ell^{BB}}{w_\ell\;C_\ell^{BB} + N_\ell^{BB}}
\end{equation}
This setup is the minimal one but has the disadvantage that one cannot change
the noise extraction while keeping the sky fixed. To allow such a possibility we
consider independent signal and noise extractions as follows. First, we
introduce new rotation matrices $ {\hat {\cal R}}_\ell $ and eigenvalues $ {\hat C}_\ell^{\pm}
$ so that now only the $ T-E $ signal covariances are diagonalized:
\begin{equation}
  \begin{pmatrix}
    {\hat C}_\ell^{TT} & {\hat C}_\ell^{TE} \\
    {\hat C}_\ell^{TE} & {\hat C}_\ell^{EE^{\phantom{H}}} \\
  \end{pmatrix} 
  =  {\hat {\cal R}}_{\ell} 
  \begin{pmatrix} 
    {\hat C}_\ell^{+} & 0 \\
    0 & {\hat C}_\ell^{-} \\
  \end{pmatrix}
   {\hat  {\cal R}}^{\rm t}_{\ell}  
\end{equation}
Then, we double all Gaussian extractions by writing the fluctuations as
\begin{equation}\label{Amps}
  A_{\ell m}^{T} = \sqrt{w_\ell}\; a_{\ell m}^{T} + \sqrt{N_\ell^{TT}}\; h_{\ell
    m}^T\;, \quad
  A_{\ell m}^{E} = \sqrt{w_\ell}\; a_{\ell m}^{E} + \sqrt{N_\ell^{EE}}\; h_{\ell
    m}^{E}\;, \quad
  A_{\ell m}^{B} = \sqrt{w_\ell}\; a_{\ell m}^{B} + \sqrt{N_\ell^{BB}}\; h_{\ell
    m}^{B}\;, \quad
\end{equation}
where $ T , E $ and $ B $ stand for temperature, $E$ polarization
and $B$ polarization, respectively, and the amplitudes $ a_{\ell m} $ are given by:
\begin{equation}
  \begin{pmatrix} 
    a_{\ell m}^{T} \\ 
    a_{\ell m}^{E^{\phantom{I}}} \\ 
  \end{pmatrix}
  = {\hat {\cal R}}_{\ell}
  \begin{pmatrix} 
    \sqrt{{\hat C}_\ell^{+}} \; g_{\ell m}^{+} \\ 
    \sqrt{{\hat C}_\ell^{-}} \; g_{\ell m}^{-} \\  
  \end{pmatrix}
 \quad , \quad a_{\ell m}^{B} = \sqrt{w_\ell\;{\hat C}_\ell^{BB}}g_{\ell m}^{B}
\end{equation}   
The new independent centered unit Gaussians $ h_{\ell m}^{Y} \; , \;
Y \equiv T, \; E, \; B $, are independent from the previous set $ g_{\ell m}^{X}  \; , \;
X\equiv +,-,B $.

\medskip

The pseudo-$C_\ell$, (that is $\overline{C}_\ell$), can now be written as
\begin{equation}
  \begin{split}
    \overline{C}_\ell^{XX} &= w_\ell\;{\tilde C}_\ell^{XX} +
    2\sqrt{w_\ell\;N^{XX}}\;Q_\ell^{XX} + N^{XX}\;P_\ell^{XX}\;,\quad X=T,E,B\\
    \overline{C}_\ell^{TE} &= w_\ell\;{\tilde C}_\ell^{TE} +
    \sqrt{w_\ell\;N^{EE}}\;Q_\ell^{TE} + \sqrt{w_\ell\;N^{TT}}\;Q_\ell^{ET} 
    + \sqrt{N^{TT}\;N^{EE}}\;P_\ell^{TE}     
  \end{split}
\end{equation}
where
\begin{equation}
  {\tilde C}_\ell^{XX'} = \frac1{2\ell+1}\sum_m a_{\ell m}^{X} \; a_{\ell m}^{X'}
  \;,\quad
  Q_\ell^{XX'} = \frac1{2\ell+1}\sum_m a_{\ell m}^{X} \; h_{\ell m}^{N_{X'}}
  \;,\quad  
  P_\ell^{XX'} = \frac1{2\ell+1}\sum_m h_{\ell m}^{X} \; h_{\ell m}^{X'}
\end{equation}
with  $ X, \; X' \equiv  T,  \; E,  \; B $. 

\medskip

This second approach allows to use the same sky and different noises, which is
needed when cumulative channels are considered to reduce noise effects. 
For instance, a cumulative channel formed by the LFI at 70 GHz and the two HFI
channels at 100 GHz and 143 GHz is obtained by simply summing
the $\chi^2$'s of Eq. (\ref{chi2}) relative to these three channels.

The above setup is based on the assumption that the noise contribution of
systematic errors is precisely assessed. If this would not be the case, bias effects
would be induced. This can be simulated in the likelihood $\chi^2$'s of
Eq. (\ref{chi2}) by using different noises in the $\overline{C}_\ell^{XX'}$ and in
the covariance built with the test multipoles $C_\ell^{XX'}$. That is, one
should make (small) variations from $N_\ell^{XX}$ to some ${N'}_\ell^{XX}$ in
Eq. (\ref{Candx}) while keeping them fixed in Eq. (\ref{Amps}) (or viceversa) and
study their impact on the parameter determination of the test cosmological
model.

\section{Forecast precision of the {\it Planck} measurements for the 
cosmological parameters without systematics}
\label{forecast_nosys}

In our Monte Carlo Markov Chains (MCMC) simulations we take as fiducial model
the $\Lambda$CDM$r$ model, that is the standard $\Lambda$CDM model augmented by
the tensor--to--scalar ratio $ r $ as described in the introduction. 
We consider MCMC simulations with both the $\Lambda$CDM$r$
and the $\Lambda$CDM$r$T model.
We denote by $\Lambda$CDM$r$T, the $\Lambda$CDM$r$ model in which
we impose the double--well inflaton potential given in 
Eq. (\ref{binon}), as described in the introduction 
and in Sect. \ref{LGtheory}.

\medskip

We consider two sets of best fit fiducial values for our parameters, as listed
in Table \ref{tab2}, where also the values of few other derived parameters are shown for
illustrative purposes. Since $ r=0 $ in the first set, the model is just the
$\Lambda$CDM model. In the second set the values $ r = 0.0427 $ and $ n_s =
0.9614 $  are chosen to lay on the theoretical curve $   r=r(n_s) $ 
dictated by the double--well inflaton potential and they correspond to the best fit value 
$ y=1.26 $ for the coupling \citep{mcmc1}
within the Ginsburg-Landau effective theory approach (see Eq. (\ref{binon})).

\medskip

We then provide estimates of the errors in the measurements of the cosmological
parameters in the following way:

\begin{itemize}

\item We produce one sky (mock data) for $ C_l^{TT}, \; C_l^{TE}, \;
    C_l^{EE} $ and $ C_l^{BB} $ from the  $\Lambda$CDM or the  $\Lambda$CDM$r$T
models (see Table \ref{tab2})     according to the procedure described in Sect. \ref{mock}.

\item We run Monte Carlo Markov Chains from this sky and obtain the marginalized
  likelihood distributions for the cosmological parameters ($ \Omega_b \, h^2 , \;
    \Omega_c \, h^2 ,\; \theta \; \tau, \; \Omega_{\Lambda} $, Age of the
    Universe, $ z_{re}, \; H_0, \; A_s,$ \; $n_s$ and $ r $)
in the two models $\Lambda$CDM$r$ and $\Lambda$CDM$r$T. To be
    precise, we study the independent $\Lambda$CDM$r$ parameters with the mock
    data produced from $\Lambda$CDM (first row of Table \ref{tab2}) and the
    independent parameters of both $\Lambda$CDM$r$ and $\Lambda$CDM$r$T with the
    mock data produced from $\Lambda$CDM$r$T (second row in Table \ref{tab2}). 
\end{itemize}

 We consider two choices for the $C_\ell-$likelihood, one without the $ B $
  modes and one with the $ B $ modes and take into account the white noise
sensitivity of {\it Planck} (LFI and HFI) in the 70, 100 and 143 GHz channels.
%\citep{PlanckBlueBook}. 
We also consider a cumulative channel whose $ \chi^2 $ is the
  sum of the $ \chi^2 $'s of the three channels above. When using different
  channels in the MCMC analysis, we use different noise realizations while
  keeping the same sky, that is the same realization of the Gaussian
  process that generated the primordial fluctuations.

  In our MCMC analysis we always take standard flat priors for the cosmological
  parameters. In particular we assume the flat priors $ 0 \leq r < 0.2 $ in the
  $\Lambda$CDM$r$ model and $ 0 \leq r< 8/60$ according to the theoretical upper limit for $ r $  
  in the $\Lambda$CDM$r$T model.

\medskip 

\begin{figure}
  \includegraphics[height=10cm]{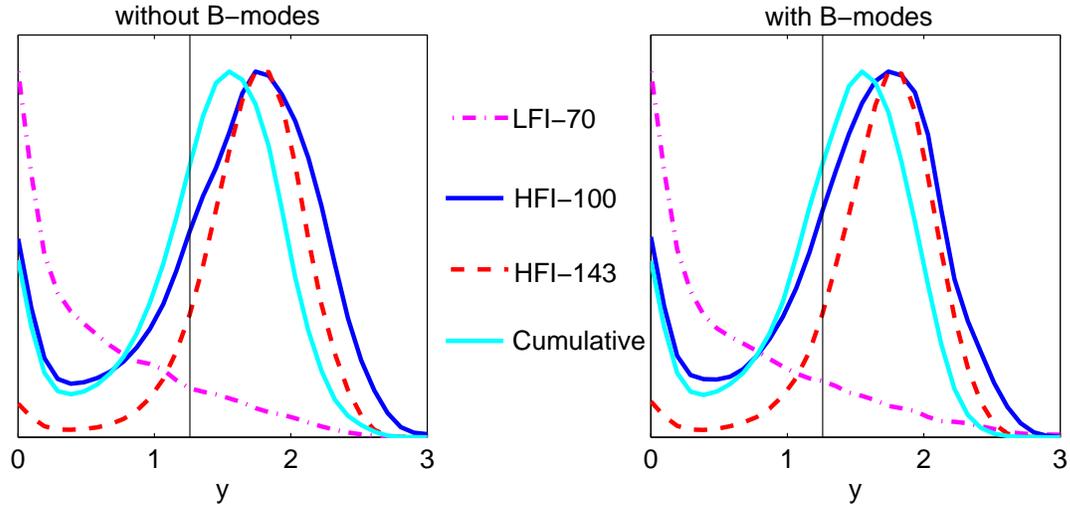}
  \caption{Marginalized likelihood distributions of the coupling constant $y$ of
    the double--well quartic inflaton potential in the $\Lambda$CDM$r$T model.
    The MCMC analysis includes (does not include) $ B $ modes in the right (left)
    panel. We display the distributions for each of the three channels and for
 the cumulative of the three channels. 
The fiducial values are indicated by a vertical thin
    black line.  The fiducial value of $ y $ is relatively well reproduced by
    the peak of the distribution in the case of the HFI-100, HFI-143 and
    cumulative channels.}
  \label{ban_y}
\end{figure}

Our findings are summarized in Figs. \ref{r0r04}-\ref{ban_y}
   where the marginalized likelihood
  distributions of the cosmological parameters are plotted for several different
  setups. In Tables \ref{tab2} to  \ref{tab4}   we list the corresponding relevant numerical values.

  Clearly, in the case of the ratio $ r $, due to the specific form of its
  likelihood distribution it is more interesting to exhibit upper and lower
  bounds rather than mean value and standard deviation as in Tables 
  \ref{tab3}  and  \ref{tab4}.  
  We report the upper bounds and, when present, the lower bounds in Tables 
  \ref{tab5}  and    \ref{tab6}.
 
Notice that best and mean values reported here for $ r $ and the
other cosmological parameters do not correspond to the real sky
but to mock MCMC generated skies as explained above.
However, the deviations between the best and the fiducial values are
relevant indicators as well as the lower and upper bounds on $r$ and the standard deviations.

The fact that the fiducial and mean values of $ r $ are very close and 
that the standard deviation $ \Delta r_{\max} $ of the distribution
of maximum values $ r_{\max} $ 
coincides with the mean value of the standard deviation of $ r $
indicate that {\it Planck} can provide measurements of high quality for $r$.

\medskip

 As expected from the relative difference in sensitivity, typically the
  distributions obtained with the HFI--100 and HFI--143 channels agree very well
  while differing markedly from those obtained with the LFI--70 channel.  Quite
  often the higher noise level in LFI--70 determines also shifts in the peak
  positions with respect to the other two channels. These shifts are within a $1\sigma$
  deviation in the LFI--70 distributions and represent therefore normal
  statistical fluctuations. The LFI--70 distribution on $ r $ when the
  fiducial value is $ r=0.0427 $ does not exhibit a peak
due to the sensitivity of this channel.

\medskip

  As expected, whenever the probability distribution for a given parameter
  is close to Gaussian, the cumulative channel produces a distribution that is
  narrower than the narrowest distribution produced by any individual
  channel. This applies to all parameters in the $\Lambda$CDM$r$ model,
  including $ r $, which has a distribution close to a left--truncated
  Gaussian for both fiducial values used. 

\medskip

  In the $\Lambda$CDM$r$T model the relation between $ n_s $ and $ r $ is
  non-linear and there are theoretical upper limits on 
$ n_s $ and $ r $ (see Fig. \ref{r04br04T}). 
  These features introduce non--Gaussianities in the distributions
  and eventually also affect any other cosmological parameter having a sensitive
  correlation with $ n_s $ and/or with $ r $, such as $ \Omega_b \, h^2 , \;  
    A_s $ or some other derived parameters. Thus, the cumulative channel provides
  some parameter distributions which are larger than those of the HFI--143
  channel, because it is affected by the LFI--70 channel, which is less sensible
  to constrain $ r $ well within its theoretical prior. This effect can be very
  well appreciated from Fig. \ref{ban_y} in which the likelihood distributions
  of the coupling constant $ y $ are plotted for the various cases considered.

\medskip

 The very limited relevance of the $ B $ modes for the 
$\Lambda$CDM$r$T model is in principle expected because the $ n_s $ value 
fixed by the $T$ modes essentially determines
$ r $ through the theoretical constraint. This property of the $ B $ modes
shows up clearly from the figures and the tables.

\subsection{Forecasts of the {\it Planck} measurements with the $\Lambda$CDM$r$ model} 

The obtained best fits, mean values and standard deviations for the 
cosmological parameters
are presented in Tables  \ref{tab3} and  \ref{tab4} for the two cases considered: without 
$B$ modes and with $B$ modes included in the $C_l$-likelihood respectively.
Each table displays the values obtained for the two test models: $\Lambda$CDM$r$
and $\Lambda$CDM$r$T. Two simulated skies (mock data) are considered: one with fiducial 
value $ r=0 $ ($\Lambda$CDM), and one with fiducial $ r=0.0427 $ 
($\Lambda$CDM$r$T). The
complete sets of fiducial values used are given in Table \ref{tab2}. All values are 
rounded to order $ 10^{-4} $ to the nearest value.

\medskip

From the Tables \ref{tab3}  to \ref{tab6} 
 we see that for the $\Lambda$CDM$r$ model the best 
result, namely a peaked distribution for $ r $,
is obtained for a non-zero fiducial value for $ r $ ($ r=0.0427 $) and with the 
$ B $ modes included in the likelihood. In this case, upper and lower bounds on $ r $ are 
obtained
$$
0.013 < r < 0.045  \quad {\rm at} \; 95\% {\rm CL}  
$$
with the best values

\medskip

$ r = 0.0594 $ and $ n_s = 0.9604 $ (without $ B $ modes);  $ r= 0.0240 $ and $ n_s = 
0.9597 $ (with $B$ modes).

\medskip

For a fiducial value $ r = 0 $, with or without $ B $ modes, the 
$\Lambda$CDM$r$ distributions peak at the value $ r= 0 $, as seen from the upper panels
of Figs. \ref{r0r04}-\ref{br0br04}. Upper bounds on $ r $ are obtained in this case. They 
result to 95\% CL
$$ 
r < 0.068 \quad {\rm without} \;  B \; {\rm modes;}  
\quad r < 0.016 \quad  {\rm with}  \; B  \; {\rm modes}
$$
with the best values

\medskip

$ r = 0.0041 $ and $ n_s = 0.9549 $ (without $ B $ modes); $ r = 0.001 $ and $ n_s = 
0.9606 $ (with $ B $ modes).

\medskip

The results on  $ n_s $ practically do not change by including or not the $B$
modes (compare the upper panels of Figs. \ref{r0r04} and \ref{br0br04}).

\medskip

The upper bound on $ r $ and the best value of $ n_s $ do not need the 
inclusion of $ B $
modes and can be obtained for fiducial $ r = 0 $. These values can be 
obtained and trusted
without including the $\Lambda$CDM$r$T model, the pure $\Lambda$CDM$r$ model
is enough to obtain them.

\medskip

We see a substantial progress in the forecasted bounds for $ r $ with 
respect to the WMAP+LSS  data set for which:
$ r < 0.20 $ in the pure $\Lambda$CDM$r$ model \citep{WMAP5,WMAP7}. 
For {\it Planck}, with the $ B $ modes 
included and a nonzero $r$-fiducial value, we get peaked distributions for $ r $ 
with a nonzero most probable value, 
lower bounds for $ r $ (and an improvement of the $ r 
$ upper bound). This is obtained by only using the $\Lambda$CDM$r$ model
alone without any input from the inflation model.
We see now in the following subsection how these forecasts can be still
considerably improved by using the $\Lambda$CDM$r$T model.

\subsection{Forecasts of the {\it Planck} measurements with the $\Lambda$CDM$r$T 
model}

With the $\Lambda$CDM$r$T model (namely when
the double--well inflaton potential is imposed), with or without $ B $ 
modes included, well peaked distributions for $ r $ are obtained together with 
upper and lower bounds and best $ r $ values. We get a considerable
gain for $ r $ with respect to the pure $\Lambda$CDM$r$ model, as can be 
seen from Tables \ref{tab3} to \ref{tab6} and Fig. \ref{r04br04T}.

The fiducial value for $ r $ ($ r = 0.0427 $) is well reproduced by the peak of the 
$\Lambda$CDM$r$T distribution both with and without $ B $ modes. 
The $\Lambda$CDM$r$T distribution
for $ r $  peaks at the non-vanishing value theoretically associated 
with the fiducial value of $ n_s $.  We get (at 95\%  CL)

\medskip

$ 0.030 < r < 0.113 $  (without $ B $ modes) ~~~~ and ~~~~  $ 0.030 < r < 
0.114 $ (with $ B $ modes),

\medskip

with the best values and 95\% CL errors

\medskip

$ r = 0.0463 \pm 0.0231 $ ~~ and ~~ $ n_s = 0.9625 \pm 0.0035 $ (without $ B $ 
modes);

\medskip

$ r= 0.0405 \pm 0.0230 $ ~~ and ~~ $ n_s = 0.9608 \pm 0.0033 $ (with $ B $ modes).

\medskip

The results with the $\Lambda$CDM$r$T model practically do not change by 
including or not the $B$
modes in the likelihood as can be seen from the figures (compare for instance 
the upper panels of Figs. \ref{r0r04} and \ref{br0br04}) and from the 
tables. This is so since the $\Lambda$CDM$r$T
model intrinsically carries a non vanishing ratio prediction, which 
shows up in agreement with the obtained marginalized distributions even 
without the inclusion of $ B $ modes.

\subsection{Conclusion}

The upper bound on $ r $ and  the best value of $ n_s $ do not require 
to include the $ B $ modes in the likelihood, and can be obtained with the 
$\Lambda$CDM$r$ model alone (see Fig. \ref{r0r04}).

\medskip

For a fiducial value $ r= 0 $, mock {\it Planck} data with the $\Lambda$CDM$r$ 
model alone, with or
without $ B $ modes in the likelihood, provide only upper bounds on $ r $ and 
most probable values
for $ n_s $.

\medskip

The same conclusions are true for a non vanishing fiducial value 
($ r = 0.0427 $) without $ B $
modes and the $\Lambda$CDM$r$ model alone.

\medskip

The inclusion of $ B $ modes for a non vanishing fiducial value 
($ r = 0.0427 $) allows peaked marginalized distributions for
$ r $ with the $\Lambda$CDM$r$ model alone and a lower bound for $ r $
(see Figs. \ref{r0r04}, \ref{br0br04} and \ref{r04br04T}).

\medskip

Lower bounds on $ r $ and most probable $ r $ values are always obtained
(with or without the $ B $ modes) for the
$\Lambda$CDM$r$T model (see Fig. \ref{r04br04T}).

\medskip

The $\Lambda$CDM$r$T model provide in all the cases, with or without the
$ B $ modes included, well peaked distributions for $ r $
on nonzero values $ r \simeq 0.04 $.

\medskip

In summary, we find that the inclusion of the theoretical model  
greatly help the recovery of the $r$ parameter. We also remark that the model is falsifiable
in the case of constraints on $n_s$ and $r$ not compatible with the banana shape of the considered 
framework.

\section{Forecast precision of the {\it Planck} measurements for the 
cosmological parameters with toy model systematics}\label{fortoy}

\subsection{Foreground residuals without bias}

We computed the cumulative marginalized likelihoods
from  the three channels including foreground residuals÷\footnote{Foreground 
residual exploited in this work
overwhelms that coming from systematics earlier discussed. 
We included the latter in some representative tests, not reported here for sake of conciseness, 
finding that, as expected, 
it does not change significantly the conclusions derived 
taking into account the foreground residual only.} 
for the cosmological parameters in the $\Lambda$CDM$r$ and the
$\Lambda$CDM$r$T models with $ B $ modes and fiducial ratios 
$ r = 0 $ and $ r = 0.0427 $. 

\medskip

The foreground residuals are introduced as an additional statistical 
error. We evaluate these statistical errors following the discussion and the toy 
model presented in Sect. \ref{toym} in which a worst case
for considering the residuals is derived (as well as a best case, and a 
intermediate or middle case for the residuals).

\medskip

\begin{figure}
 \includegraphics[height=15cm,angle=90]{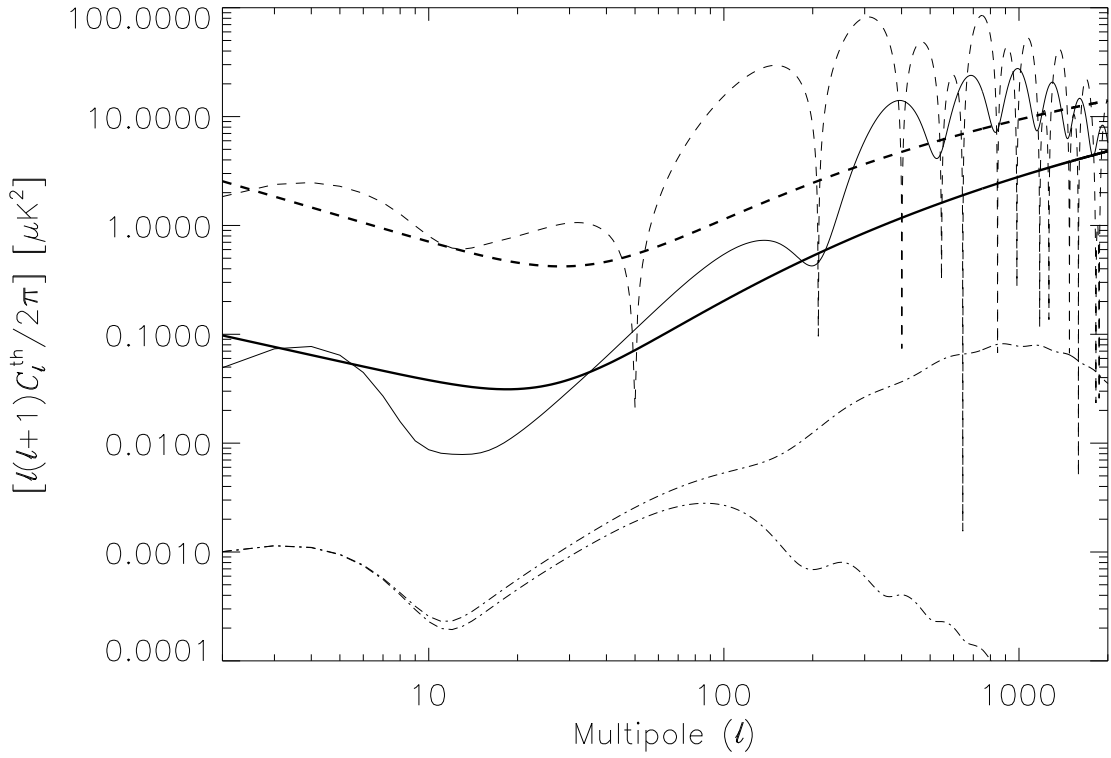}
\caption{Comparison between our model of foreground residual in the $E$ mode 
(solid line) and in the $TE$ mode (dashes) with 
a typical CMB angular power spectrum for the same modes and our fiducial B mode
(i.e. with $ r = 0.0427 $) including (upper dashed-dot line) or not
(lower dashed-dot line) the contribution by lensing.
Foreground residuals in the $ B $ modes are assumed equal to those
in the $ E $ modes.}
  \label{foreres_TE_EE}
\end{figure}

\begin{figure}
  \includegraphics[height=7cm]{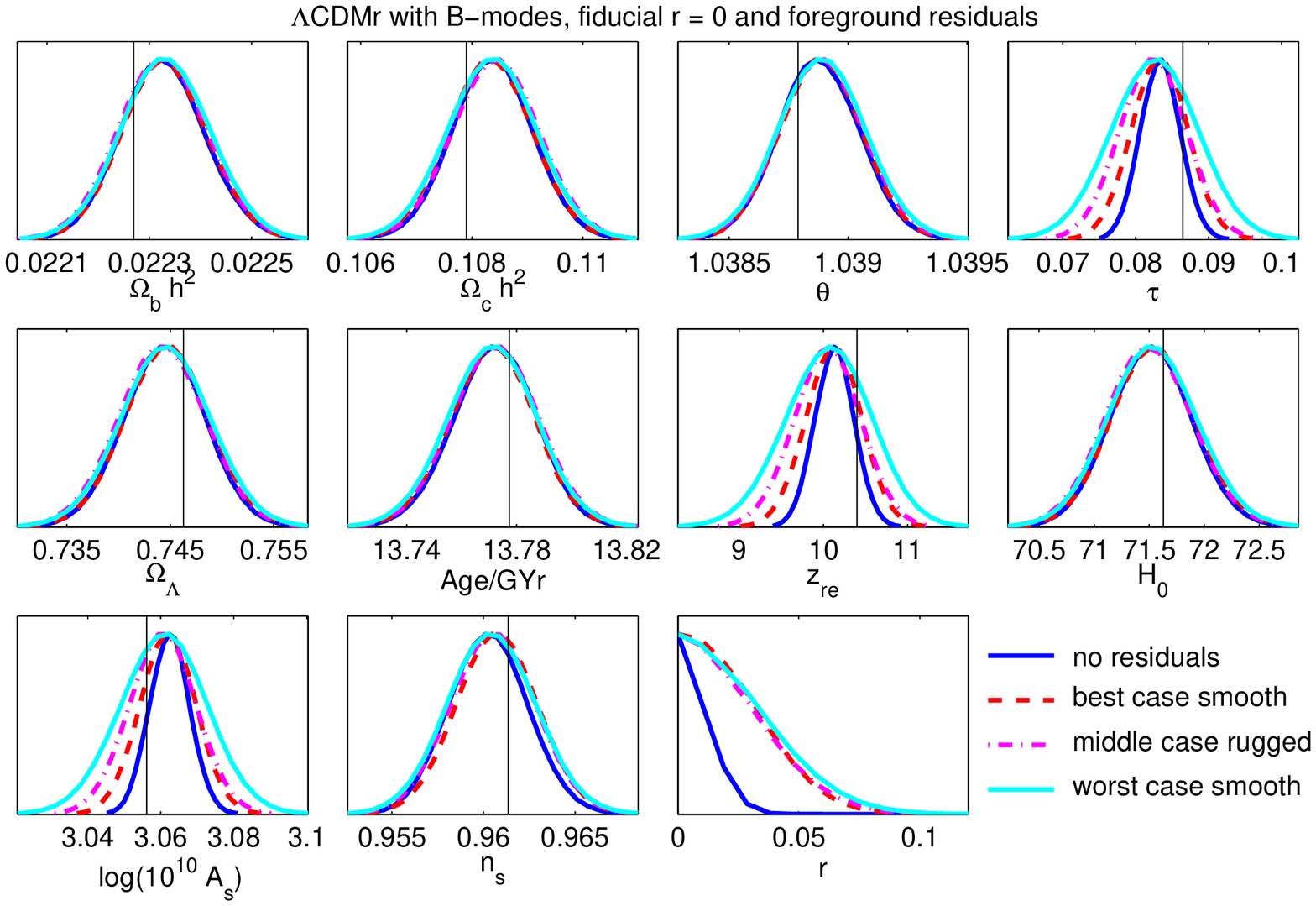} \\
  \includegraphics[height=7cm]{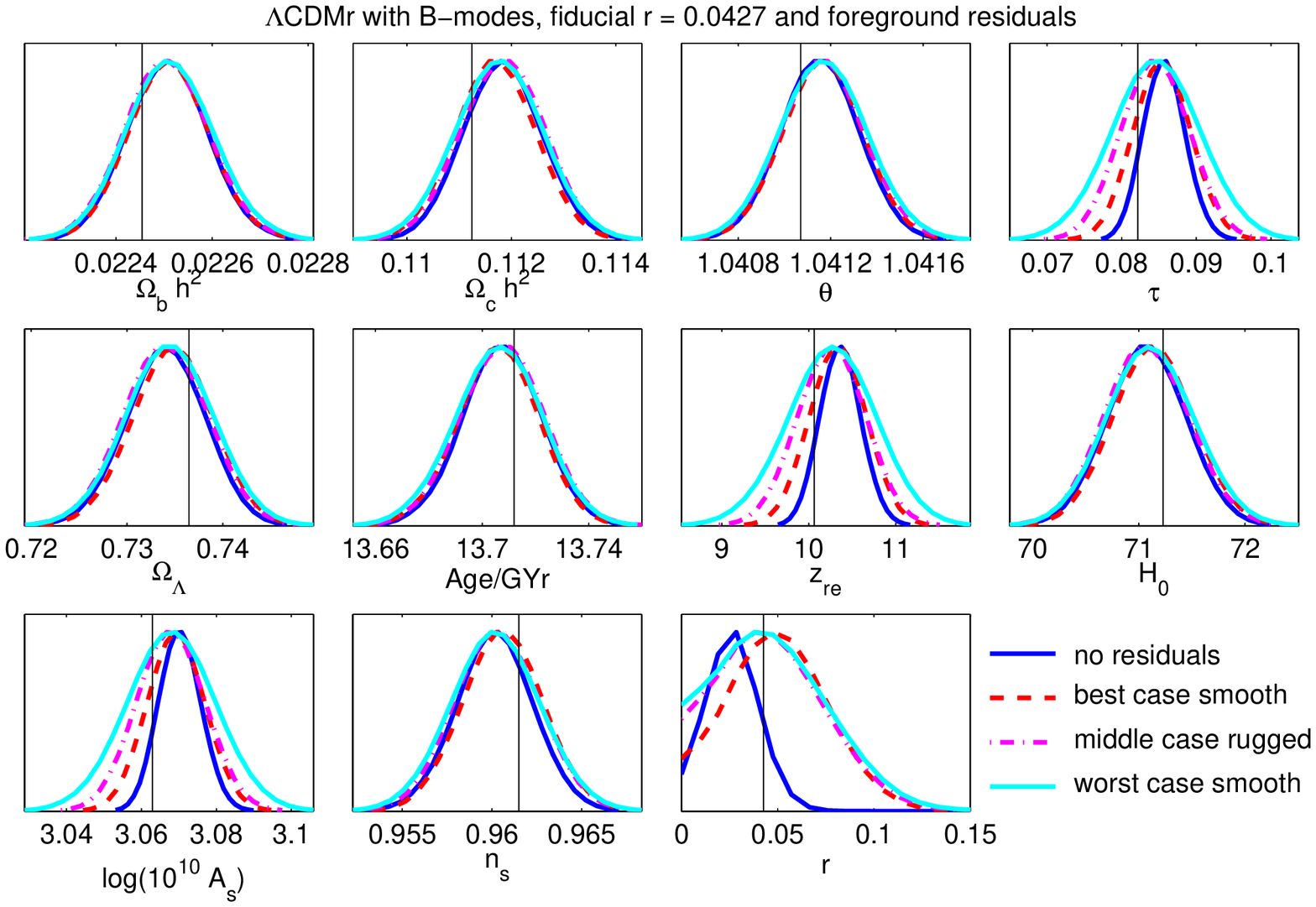}
  \caption{Cumulative $3-$channel marginalized likelihood distributions,
    including $ B $ modes and foreground residuals, of the cosmological
parameters for the $\Lambda$CDM$r$ model.  The fiducial ratio is $ r = 0 $
    in the upper panel and $ r = 0.0427 $ in the lower. We plot the
distributions  in four cases: (a) without residuals, (b) with 30\% of the
toy model residuals in the $TE$ and $E$ modes displayed in Fig. \ref{foreres_TE_EE}
and $ 16 \mu K^2 $ in the 
$T$ modes, (c) with the toy model residuals in the $TE$ and $E$ modes 
displayed in Fig. \ref{foreres_TE_EE} and 
$ 160 \mu K^2 $ in the $T$ modes, (d) with 65\% of the toy model 
residuals in the $TE$ and $E$ modes displayed in Fig. \ref{foreres_TE_EE}
and $ 88 \mu K^2 $ in the $T$ modes rugged by 
Gaussian fluctuations of $ 30 \% $ relative strength. }
  \label{resbr0br04}
\end{figure}

\begin{figure}
  \includegraphics[height=10cm]{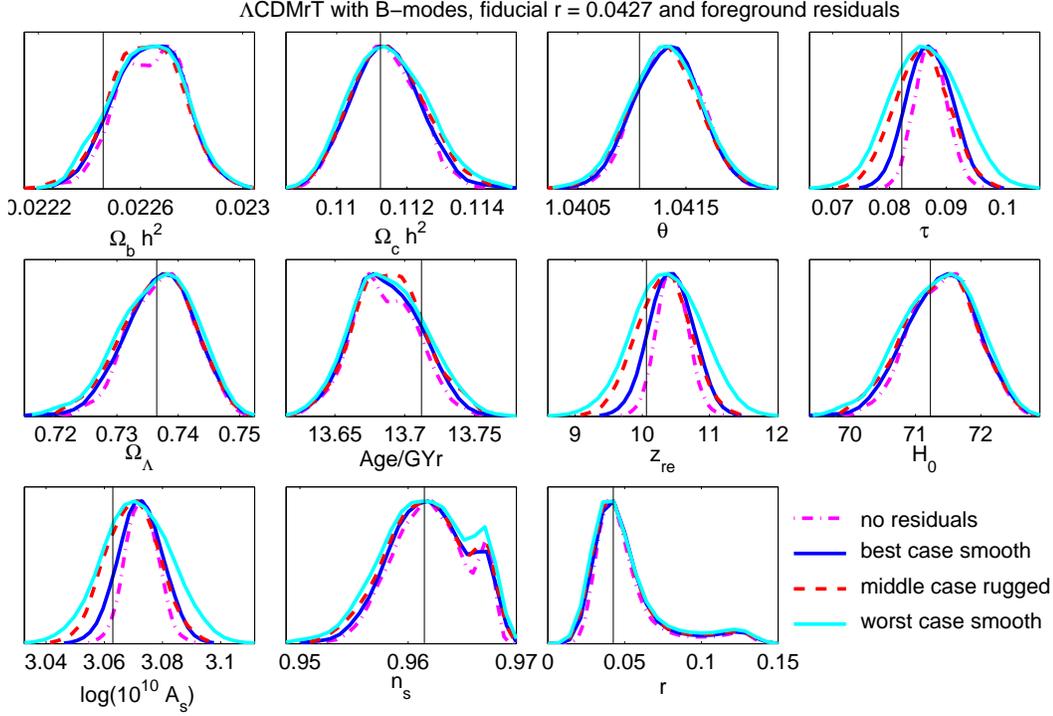}
  \caption{Cumulative marginalized likelihoods from the three channels for the
    cosmological parameters for the $\Lambda$CDM$r$T model including $ B $ modes
    and fiducial ratio $ r = 0.0427 $ and the foreground residuals.  We plot
    the cumulative likelihoods in four cases: (a) without residuals, (b) with
    0.3 of the worst case residuals in the $TE$ and $E$ modes and $ 16 \mu K^2 $ in
    the $T$ modes, (c) with the worst case residuals in the $TE$ and $E$ modes and $
    160 \mu K^2 $ in the $T$ modes, (d) 
    with 65\% of the toy model 
residuals in the $TE$ and $E$ modes displayed in Fig. \ref{foreres_TE_EE}
and $ 88 \mu K^2 $ in the $T$ modes rugged by 
Gaussian fluctuations of $ 30 \% $ relative strength.}
  \label{resbr04T}
\end{figure}

We plot the likelihoods for the cumulative of the three channels
in four cases (see Figs. \ref{resbr0br04}-\ref{resbr04T}):
\begin{itemize}
\item (a) without residuals;
\item (b) with 30\% of the toy model
residuals in the $TE$ and $E$ modes displayed in 
Fig. \ref{foreres_TE_EE} and $ 16 \, \mu K^2 $ in the $T$ modes;
\item (c) with the toy model residuals in the $TE$ and $E$ modes displayed 
in Fig. \ref{foreres_TE_EE} and $ 160 \, \mu K^2 $ in the $T$ modes;
\item (d) with 65\% of the toy model
residuals in the $TE$ and $E$ modes displayed in Fig. \ref{foreres_TE_EE} and $ 88 \, \mu K^2 $ in the 
$T$ modes rugged by Gaussian fluctuations of $ 30 \% $ relative strength.
\end{itemize}

\medskip

The likelihood distributions with and without $ B $ modes result almost the 
same when including the residuals. Only the cosmological parameters 
sensitive to the $ B $ modes do appear to be affected by the residuals, namely, 
$ \tau, \; z_{re} $ and $ r $. This is so in the
$\Lambda$CDM$r$ model in which for a fiducial value $ r = 0 $, the 
upper bound in $ r $ does change by including the residuals
(see Fig. \ref{resbr0br04}). This change is smaller for a fiducial ratio $ 
r = 0.0427 $.
In this case, the presence of a lower bound for $ r $ (at 68\% CL) 
does remain even by including the residuals. The lower bound remains at 
95 \% CL in the best case smooth residuals.
The main numbers are displayed in Tables \ref{tab7} to  \ref{tab9}.

\medskip

The $ r $ distribution for fiducial ratio $ r = 0.0427 $ is much
clearly peaked on a value of  $ r $ near the fiducial one in the 
$\Lambda$CDM$r$T model than in the $\Lambda$CDM$r$ model 
(compare Figs. \ref{resbr0br04} and \ref{resbr04T}).
In any case, the best value for $ r $ in the presence of residuals
is about $ r \simeq 0.04 $ (near the fiducial value) both for the 
$\Lambda$CDM$r$ and the $\Lambda$CDM$r$T models.

The $\Lambda$CDM$r$T model turns to be robust, it is very stable with 
respect to the inclusion of the residuals as
its marginalized likelihood 
distributions do not change (and we have seen in Sect. \ref{mock}
that they do not change neither
with respect to the inclusion or not of $B$ modes). The main numbers are 
included in Tables \ref{tab7} to \ref{tab9}. We see in the $\Lambda$CDM$r$T model
that we have again for $ r $  at 95 \%  CL:
$$ 
0.028 < r < 0.116 \quad {\rm with ~the ~best ~value} \quad r = 0.04 \; .
$$

In summary, foreground residuals only affect $ B $ modes and therefore 
only the cosmological parameters sensitive to  $ B $ modes are affected.

\subsection{Probability to detect $ r $ from $ B $ modes in the HFI-143 channel}\label{143}

\begin{figure}
  \includegraphics[height=10cm]{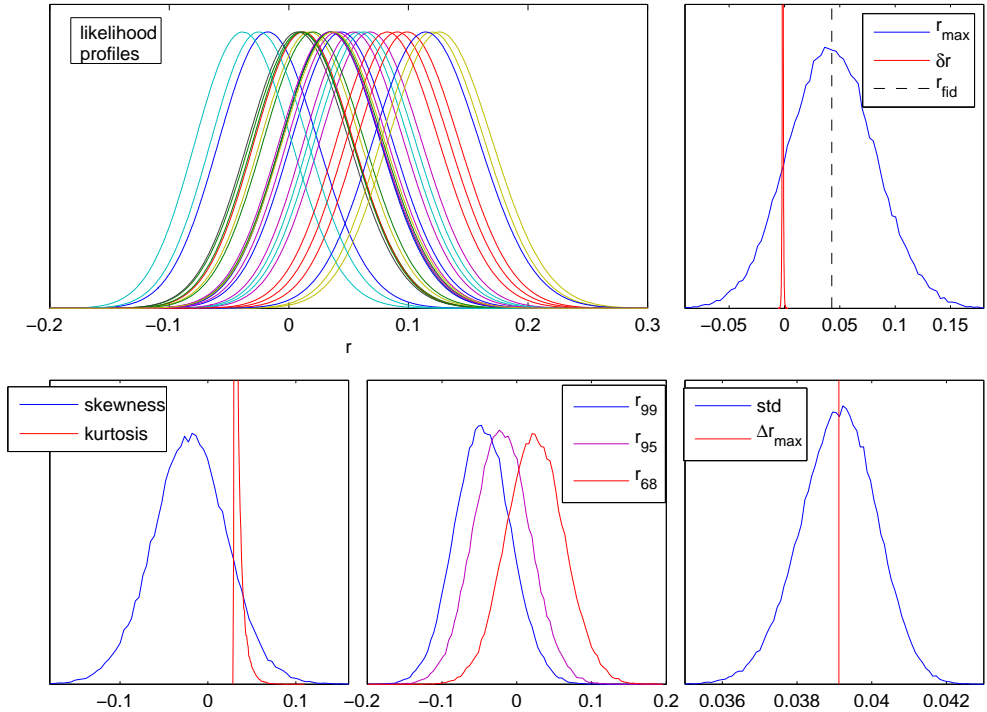}
  \caption{Upper left panel: the likelihood profiles for the different skies. 
  Upper right panel:
 $ r_{max} $ and  $ \delta  r_{max} \equiv r_{max} - r_{mean} $ 
and the fiducial $ r , \; r_{fid} $. 
Lower left panel: the skewness and the kurtosis.
Lower middle panel: the 99\% CL, 95\% CL, and 68\% CL lower bounds
for $ r : \; r_{99}, \; r_{95} $ and $ r_{68} $. Lower right panel: the standard 
deviation std of the $ r $ distributions for each sky and the standard 
deviation $ \Delta r_{max} $ of the $ r_{max} $ distribution.
These results correspond to a level of foreground residual equal 
to 30\% of the toy model displayed in Fig. \ref{foreres_TE_EE}. 
The sensitivity of the 143 GHz channel
is exploited here.}
  \label{probd}
\end{figure}

\begin{figure}
\includegraphics[height=10cm]{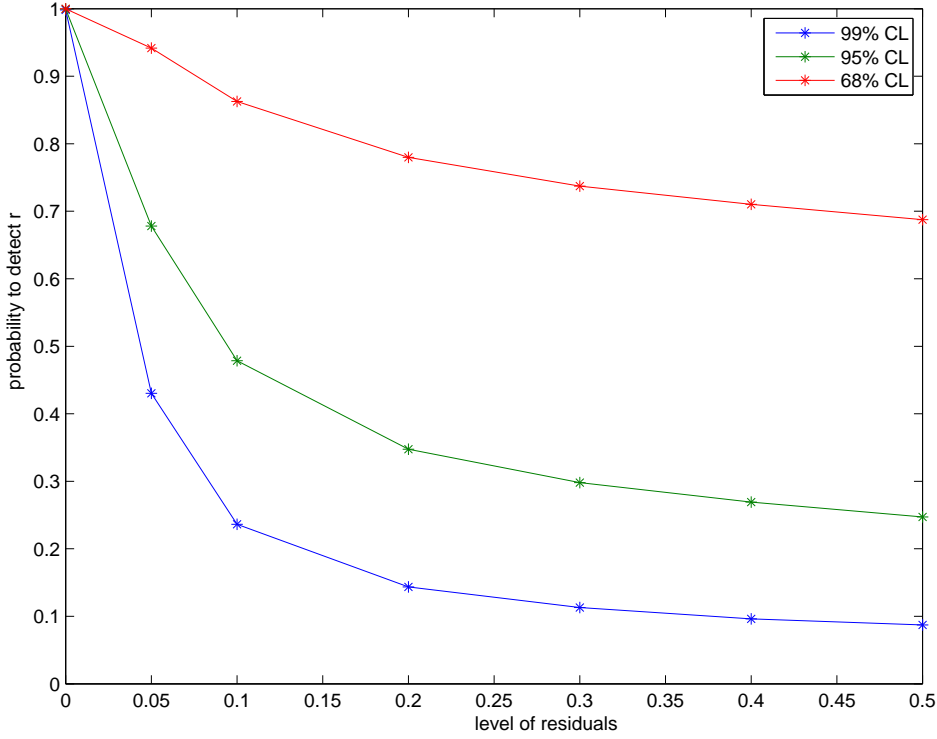}
\caption{99\% CL, 95\% CL and 68\% CL lower bounds for $ r $ 
as functions of the fraction of foreground residual of the worst case.
For 30\% foreground residual case only a 68\% CL 
detection is very likely. For a 95\% CL detection 
the level of foreground residual should be reduced to 10\% or 
lower of the toy model displayed in Fig. \ref{foreres_TE_EE}. 
The sensitivity of the 143 GHz channel is exploited here.}
\label{probd2}
\end{figure}

Figs. \ref{probd} and \ref{probd2} describe the probability of
detection of $ r $ from $B$ modes in the HFI-143 channel.

In order to assess the probability for {\it Planck} to detect $ r $ we
consider the $B$ mode detection by the most sensitive HFI-143 channel.
We follow the following method:

\begin{itemize}
\item{We extract $10^5$ skies obtaining the corresponding multipoles $ A_{lm} $
from the $\Lambda$CDM$r$T model according to the procedure described 
in Sect. \ref{mock}. We choose $ r=0.0427 $ as fiducial value.}
\item{We compute all the corresponding likelihood profiles only for $ r $.
That is, freezing out all the other parameters to their fiducial values.}
\item{We compute the interesting properties of each likelihood profile, like
the most likely value $ r_{max} $, the mean value $ r_{mean} $, the standard 
deviation  $ \Delta r_{max} $ of the $ r_{max} $ distribution, the skewness and the kurtosis. 
This last measures the departure from a Gaussian  likelihood.}
\item{We finally compute the 99\% CL, 95\% CL, and 68\% CL lower bounds
for $ r : \; r_{99}, \; r_{95} $ and $ r_{68} $}.
\end{itemize}
In the five panels of Fig. \ref{probd} we plot the likelihood profiles
for the different skies, $ r_{max} $ and  $ \delta r_{max} \equiv r_{max} - r_{mean} $.
Notice that $ r_{max} $ is an unbiased estimator of the true value, 
since its expectation value $ r_{mean} $
throughout many skies coincides with the fiducial $ r $.
$ \Delta r_{\max} $ is the standard deviation of the $ r_{max} $  distribution. 
We find that $ \Delta r_{\max} $ always
agrees extremely well with the mean value of the standard deviation
of $ r $ in each likelihood profile for each different sky. This fact means that 
asymptotically for a large number of skies the 
width of the $ r $ profile is an unbiased estimator of the actual uncertainty in $ r $.

All these results were obtained for a level of foreground residual equal to 
30\% of the toy model displayed in Fig. \ref{foreres_TE_EE}.

We plot in Fig. \ref{probd2} 
the 99\% CL 95\% CL, 68\% CL lower bounds for $ r $: $ r_{99}, \;  r_{95} $ and
$\; r_{68} $, respectively as functions of the fraction of foreground residual
of the worst case. These lower bounds are consistent, since
they fail more or less 99\%, 95\%, 68\% of the sky extractions. This last property
is true only if the prior $ r>0 $ is not enforced in the likelihood.
That is why we get a non-zero likelihood on negative values of $ r $. Of course,
only positive values of $ r $ are meaningful and this allows us to
 define the probability of detection of $ r $, to  99\% CL, 95\% CL and 
68\% CL, as the fraction of skies which gives positive 99\% CL, 95\% CL
and 68\% CL lower bounds, respectively.

These probabilities of detection are displayed in Fig. \ref{probd2}.
At the level of foreground residual equal to 30\% of the considered toy model 
only a 68\% CL (one sigma)
detection is very likely. For a 95\% CL detection (two sigmas)
the level of foreground residual should be reduced to 10\% or lower of the 
toy model displayed in Fig. \ref{foreres_TE_EE}.

\medskip

Lensing was not considered in the analysis on $r-$detection probability.
Residuals and lensing affect the detection of B-modes
in complementary ways, with the effect of residuals stronger than 
that of lensing. As a consequence, lensing plus residuals
can spoil the detection of $ r $ even when residuals are assumed
at the 30\% level of the toy model displayed in Fig. \ref{foreres_TE_EE}. 
On the contrary, lensing in the absence
of residuals still allows a detection of $ r $. For example, 
several MCMC simulations show that our lower
bounds on $ r $ are not significantly affected by lensing in the absence 
of foreground residuals. It should be clear that if the theoretical 
constraint $ r = r(n_s) $ of the $\Lambda$CDM$r$T model is imposed 
on the MCMC analysis, $ r $ has always well defined lower bounds 
regardless of lensing and/or residuals.

The forecasted probability of detecting $ r $ 
is based on the statistics of the
shape of the $r$ -likelihood. This shape determines whether a detection of
$ r $ can be claimed with a given confidence level.
But real CMB experiments can observe only one realization (only
one sample): the actual observed sky from which a single likelihood for the
B-mode multipoles  (and hence the value of $ r $)  is derived. So, the
possibility of correctly inferring the value of $ r $ from one single
(albeit very large) sample depends heavily on the sample itself, and
therefore, in view of the detection probability found here,
whether $ r $ will be or will be not detected depends also of a question
on luck.

\medskip

In addition, the results for many skies presented in this section show
the consistency of our whole approach to determine $ r $.
Similar results are valid for the other cosmological parameters.

\subsection{Bias effect in the foreground residuals}
\label{efbias}

\begin{figure}
\includegraphics[height=10cm]{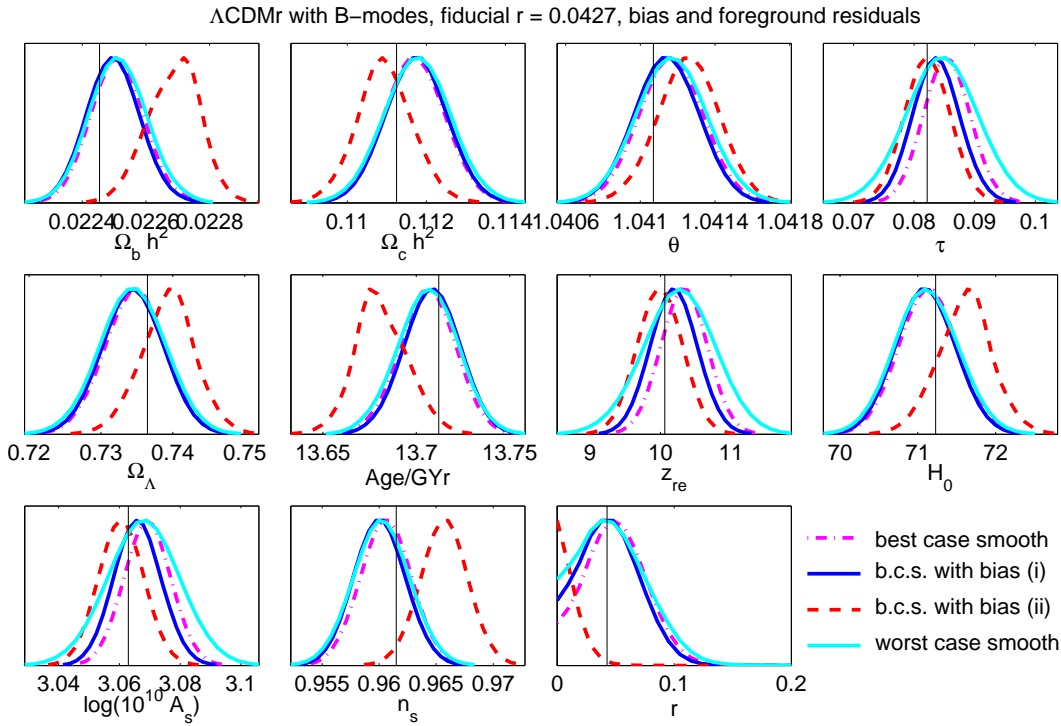}
\caption{Likelihoods profiles with bias and foreground residuals including
$ B $ modes for the $\Lambda$CDM$r$ model and fiducial value $ r=0.0427 $.
We plot the best and worst smooth cases of the residuals without bias
and  the best smooth case for the residuals including the bias according 
to the cases (i) and (ii) in Sect. \ref{efbias}, namely small and large bias 
cases respectively. }
\label{probd3}
\end{figure}

Here we consider two extremal cases of bias, modelled as being an imprecise determination
of the foreground residuals $ R_\ell $. We keep fixed the
residuals introduced in the noise of the test covariances, while we change the residuals
$ R_\ell^{XX} $ to some $ {R'}_\ell^{XX} $ in the the noise of the observations, that is in
Eq. (\ref{covres}). Then, we write
\be \label{bias1}
  {R'}_\ell^{XX} = R_\ell^{XX} \; \left(1 + \beta_\ell^X \right) ~~,~~ X= T, E, B \; ,
\ee
where for the numbers $ \beta_\ell $ we consider the two extremal cases:
\begin{itemize}
\item(i) Independent flat random numbers $ \beta_\ell $ from $ -0.5 $ to $ 0.5 $.
Since in this case $ \beta_\ell $ fluctuates randomly around zero the
effect of the bias mostly cancel out and the cosmological parameters suffer little change as 
depicted in Fig. \ref{probd3}.
\item(ii) Uniform ramps from $ -a^X $ to $ a^X $ as $ \ell $ varies from 2 to 2100,
 $ a^X $ varying randomly up to a 20\%  around 0.5 with $ X=T, \; E, \; B $.
This means to choose
\be \label{caso2}
\beta_\ell^X = a^X \; \left(\frac{\ell - 1051}{1049} \right) \; .
\ee
Notice that there is always a non-zero value for $ a^X $ despite its fluctuations
and therefore there is a significant bias effect over the modes.
In this case the bias depresses the estimated multipoles at
low $ \ell $ and increases them at high $ \ell $. Thereby increasing the expected 
values of $ n_s $ and depressing those of $ r $. We see from Fig. \ref{probd3}
%that the detection of $ r $ is practically spoiled in this case.
that $ r $ is not anymore detected in this case despite its fiducial value is always $ r=0.0427 $.
\end{itemize}
We consider a level of 30\% of the toy model of foreground residuals 
displayed in Fig. \ref{foreres_TE_EE} for the bias effect.
In the case we change the overall sign of $ \beta_\ell^X $
in Eq. (\ref{caso2}), we introduce an additional spurious power in the estimation at
low $ \ell $ and a depression of the power at high $ \ell $. 
This would erroneously increase the probability to detect $r$. 

The peaks in the cosmological parameters (with the exception of $ r $) get shifted 
mainly due to the bias from the $T$ modes. However, they stay within one or two $ \sigma $ of the
WMAP values.

In case we do not add bias in the $T$ modes, only $ r $ is affected significantly
by the bias. Namely, in case (ii) without bias in $T$ modes, all cosmological parameters 
except $ r $ peak practically at the same value as in absence of bias. On the contrary,
the likelihood distribution for $ r $ is determined by the bias on the $B$ modes
and turns to be similar to the one in Fig. \ref{probd3} for the bias case (ii).

%Fig.? \ref{probd3} where we plot the likelihood
%distributions including the two considered types of biais.

The bias introduced in case (ii) goes in the opposite direction to the 
theoretical double--well models where $ n_s $ increases with $ r $ 
(see Fig. \ref{banana}). 

We only present here the bias for the $\Lambda$CDM$r$ model. The likelihood
distributions for the $\Lambda$CDM$r$T model including bias are similar to those
of the  $\Lambda$CDM$r$  model except for $ r $ 
where a lower bound shows up due to the theoretical constraint.

We only consider here two extreme cases of bias: case (i) where bias is practically
harmless and case (ii) where it distorts significantly the cosmological parameters,
especially $ r $ which is not anymore detected.

\section{Final Conclusion}

In this paper we provide a precise forecast for the {\it Planck} results
on cosmological parameters, in particular for the tensor--to--scalar
ratio $ r $. These new forecasts go far beyond the published ones (see e.g.
\citet{PlanckBlueBook}, \citet{otros})
and pave the road for a promising scientific exploitation and 
interpretation of the {\it Planck} data (once cleaned from the different
astrophysical foregrounds). 

We appropriately combined the following, as main ingredients:
the current public available knowledge of {\it Planck} instrument sensitivity and a reasonable toy model 
estimation of the residuals from systematic errors and foregrounds;
the highly predictive theory setup \citep{reviu,mcmc1,mcmc2} provided by the Ginzburg-Landau
approach to inflation to produce and analyze the skies (mock data)
which allows a decisive gain in the physical insight and
data analysis;
precise MCMC methods to produce the skies (mock data)
and to analyze them.
This turns into an improvement in the physical analysis, in particular for the ratio $ r $. 

It must be also
stressed that, in the considered framework, better measurements for $ n_s $ will improve
the predictions on $ r $ from the $T$, $TE$ and $E$ modes even if a
secure detection of $B$ modes will be still lacking.
We remark also that the model is falsifiable
in the case of constraints on $ n_s $ and $ r $ not compatible with the banana shape 
of the considered framework.

The lower bounds and most probable value inferred from WMAP for  $ r $ 
($ r \simeq 0.04 $) in the considered framework support
the search for $B$ mode polarization in {\it Planck} data and the future 
CMB  $B$ oriented polarization missions.


%% Included in this acknowledgments section are examples of the
%% AASTeX hypertext markup commands. Use \url without the optional [HREF]
%% argument when you want to print the url directly in the text. Otherwise,
%% use either \url or \anchor, with the HREF as the first argument and the
%% text to be printed in the second.

\acknowledgments

We thank Maria Cristina Falvella for her invaluable help and useful 
stimulating discussions. We thank an anonymous referee for his/her constructive comments.
We acknowledge the use of the Legacy Archive for Microwave Background Data Analysis (LAMBDA). 
Support for LAMBDA is provided by the NASA Office of Space Science. 
Some of the results in this paper have been derived using the HEALPix 
\citep{2005ApJ...622..759G} package. 
This work has been done in the framework of the {\it Planck} LFI activities. 
We acknowledge the support by ASI through 
ASI/INAF Agreement I/072/09/0 for the {\it Planck} LFI Activity of Phase E2 and I/016/07/0 COFIS.
The Monte Carlo simulations were performed on the Turing cluster
  of the Dipartimento di Fisica ``G. Occhialini'', Universit\`a
  Milano-Bicocca.

\clearpage

\begin{table}
	\begin{tabular}{lccc}
\hline
	Frequency channel &	$143\,$GHz	& $100\,$GHz	& $70\,$GHz \\
\hline
%	InP detector technology	& MIC	& MIC	& MMIC \\
	Angular resolution [arcmin]	& 7.1 &	9.5 &	13 \\
	$\delta$T per FWHM$^2$ pixel [$\mu$K] & 4.2	 & 4.8 & 17.2 \\
	$\delta$Q, $\delta$U per FWHM$^2$ pixel [$\mu$K] & 8.1 & 7.7  & 24.3 \\
%	$\delta$T/T per pixel [$\mu$K/K]	& 2.67    & 3.67 & 6.29 \\
%	$\delta$T/T per pixel [$\mu$K/K]	& 2.6    & 3.6	& 6.2 \\
%	Number of radiometers (or feeds)	& 4 (2)	& 6 (3)	& 12 (6) \\
%	Effective bandwidth [GHz]	& 6	&8.8	& 14 \\
%	System noise temperature [K]	& 10.7	& 16.6	& 29.2 \\
%	White noise per channel [$\mu$K $\cdot \sqrt{{\rm s}}$] & 116 & 113 & 105 \\
%	Systematic effects [$\mu$K]	& $<$ 3	& $<$ 3	& $<$ 3 \\
\hline
\end{tabular}
\caption{{\it Planck} performance in the three frequency channels exploited in this work.
The average sensitivity per FWHM$^2$ resolution element 
($\delta$T, $\delta$Q, $\delta$U) is given in
CMB temperature units (i.e. equivalent thermodynamic temperature) for 28 
months of integration, almost corresponding to four sky surveys.}
\label{table:sens}
\end{table}

\begin{table}
%\tabletypesize{\scriptsize}
%\rotate
%\tabletypesize
%\small{8pt}
  \centering
\begin{tabular}{|p{1.8cm}|p{1cm}|p{1cm}|p{1cm}|p{1cm}|p{2cm}|p{1cm}|p{1cm}||p{1cm}|p{1cm}|p{1cm}|}
%\begin{tabular}{|c|c|c|c|c|c|c|c||c|c|c|}
    \hline
    &~~$\Omega_b \, h^2$~~&~~$\Omega_c \, h^2$~~&~~~~$\theta$~~~~&~~~~$\tau$~~~~ 
    &$\log(10^{10}A_s)$&~~~$n_s$~~~&~~~~$r$~~~~&~~$\Omega_\Lambda$~~&~~~$H_0$~~~&~~$z_{re}$~~\\
    \hline \hline
    $\Lambda$CDM & $0.0223$ & $0.1079$ & $1.0387$ & $0.0864$ 
    & $3.0561$ & $0.9613$ & $0$ & $0.7463$ & $71.628$ & $10.399$\\
    \hline
    ~$\Lambda$CDM$r$T~ & $0.0224$ & $0.1112$ & $1.0410$ & $0.0821$ 
    & $3.0629$ & $0.9615$ & $0.0427$ & $0.7364$ & $71.228$ & $10.062$ \\
    \hline
  \end{tabular}
  \caption{Fiducial parameters for the two considered models: the standard $\Lambda$CDM model
    and the $\Lambda$CDM$r$T model. In the $\Lambda$CDM$r$T model
    we constrain $ r=r(n_s) $ by the double--well inflaton potential given in Eq. (\ref{binon}) 
    as depicted on the upper border of the banana-shaped region Fig. \ref{banana}.}
\label{tab2}
\end{table}

\begin{table}
  \centering
  \begin{tabular}{|p{1.8cm}||p{1cm}|p{1cm}|p{1cm}||p{1cm}|p{1cm}|p{1cm}||p{1cm}||p{1cm}|p{1cm}|p{1cm}|}
%  \begin{tabular}{|c||c|c|c||c|c|c||c|c|c|}
    \hline
    Sky data & \multicolumn{3}{c||}{$\Lambda$CDM: fiducial $ r=0 $} &
    \multicolumn{6}{c|}{$\Lambda$CDM$r$T: fiducial $ r=0.0427 $}       \\ \hline
    test model&\multicolumn{3}{c||}{$\Lambda$CDM$r$}
    &\multicolumn{3}{c||}{$\Lambda$CDM$r$ }
    &\multicolumn{3}{c|}{$\Lambda$CDM$r$T } \\ \hline 
    without $B$ modes& ~~best~~ & ~~mean~~ & stddev 
    & ~~best~~ & ~~mean~~ & stddev & ~~best~~ & ~~mean~~ & stddev \\ \hline
    $\Omega_bh^2$        & $0.0223$ & $0.0223$ & $0.0001$ & $0.0225$ & $0.0225$ 
    & $0.0001$ & $0.0227$ & $0.0226$ & $0.0001$\\ \hline
    $\Omega_ch^2$        & $0.1085$ & $0.1082$ & $0.0007$ & $0.1118$ & $0.1116$
    & $0.0007$ & $0.1112$ & $0.1114$ & $0.0010$\\ \hline
    $\theta$             & $1.0389$ & $1.0388$ & $0.0002$ & $1.0411$ & $1.0412$
    & $0.0002$ & $1.0415$ & $1.0413$ & $0.0003$\\ \hline
    $\tau$               & $0.0832$ & $0.0833$ & $0.0027$ & $0.0858$ & $0.0857$
    & $0.0027$ & $0.0871$ & $0.0872$ & $0.0028$\\ \hline
    $\log[10^{10} A_s]$  & $3.0479$ & $3.0618$ & $0.0054$ & $3.0703$ & $3.0697$
    &$0.0056$ & $3.0722$ & $3.0729$ & $0.0055$\\ \hline
    $n_s$                & $0.9549$ & $0.9609$ & $0.0021$ & $0.9604$ & $0.9607$ 
    & $0.0022$ & $0.9625$ & $0.9619$ & $0.0035$\\ \hline
    $r$                  & $0.0041$ & $0.0284$ & $0.0206$ & $0.0594$ & $0.0532$ 
    & $0.0277$ & $0.0463$ & $0.0510$ & $0.0231$\\ \hline \hline
    $\Omega_\Lambda h^2$ & $0.7436$ & $0.7451$ & $0.0038$ & $0.7344$ & $0.7351$ 
    & $0.0040$ & $0.7388$ & $0.7369$ & $0.0059$\\ \hline
    $H_0$                & $71.437$ & $71.577$ & $0.3610$ & $71.111$ & $71.165$ 
    & $0.3638$ & $71.591$ & $71.397$ & $0.5288$\\ \hline
    $z_{\rm re}$         & $10.128$ & $10.121$ & $0.2267$ & $10.375$ & $10.358$ 
    & $0.2328$ & $10.414$ & $10.446$ & $0.2276$\\ \hline
  \end{tabular}            
  \caption{Best fits, mean values and standard deviations for cosmological
    parameters when $ B $ modes are not included in the $ C_\ell$-likelihood. All values
    are rounded to order $ 10^{-4} $ to the nearest value and correspond
    to the cumulative channel whose $ \chi^2 $ is the
    sum of the $ \chi^2 $'s of the three channels HFI-100, HFI-143 and LFI-70.}
\label{tab3}
\end{table}

\begin{table}
  \centering
   \begin{tabular}{|p{1.8cm}||p{1cm}|p{1cm}|p{1cm}||p{1cm}|p{1cm}|p{1cm}||p{1cm}||p{1cm}|p{1cm}|p{1cm}|}
%  \begin{tabular}{|c||c|c|c||c|c|c||c|c|c|}
    \hline
    Sky data & \multicolumn{3}{c||}{$\Lambda$CDM: fiducial $r=0$} &
    \multicolumn{6}{c|}{$\Lambda$CDM$r$T: fiducial $r=0.0427$}       \\ \hline
    test model&\multicolumn{3}{c||}{$\Lambda$CDM$r$}
    &\multicolumn{3}{c||}{$\Lambda$CDM$r$ }
    &\multicolumn{3}{c|}{$\Lambda$CDM$r$T } \\ \hline 
    with $B$ modes& ~~best~~ & ~~mean~~ & stddev 
    & ~~best~~ & ~~mean~~ & stddev & ~~best~~ & ~~mean~~ & stddev \\ \hline
    $\Omega_bh^2$       & $0.0223$ & $0.0223$ & $0.0001$ & $0.0225$ & $0.0225$ 
    & $0.0001$ & $0.0226$ & $0.0226$ & $0.0001$\\ \hline
    $\Omega_ch^2$       & $0.1081$ & $0.1084$ & $0.0007$ & $0.1118$ & $0.1118$
    & $0.0007$ & $0.1117$ & $0.1114$ & $0.0010$\\ \hline
    $\theta$            & $1.0389$ & $1.0389$ & $0.0002$ & $1.0412$ & $1.0411$
    & $0.0002$ & $1.0412$ & $1.0413$ & $0.0003$\\ \hline
    $\tau$              & $0.0834$ & $0.0840$ & $0.0027$ & $0.0866$ & $0.0857$
    & $0.0027$ & $0.0865$ & $0.0873$ & $0.0028$\\ \hline
    $\log[10^{10} A_s]$ & $3.0620$ & $3.0624$ & $0.0053$ & $3.0721$ & $3.0703$ 
    & $0.0054$ & $3.0720$ & $3.0728$ & $0.0056$\\ \hline
    $n_s$               & $0.9606$ & $0.9603$ & $0.0021$ & $0.9597$ & $0.9602$ 
    & $0.0021$ & $0.9608$ & $0.9621$ & $0.0033$\\ \hline
    $r$                 & $0.0010$ & $0.0060$ & $0.0050$ & $0.0240$ & $0.0275$ 
    & $0.0096$ & $0.0405$ & $0.0516$ & $0.0230$\\ \hline \hline
    $\Omega_\Lambda h^2$& $0.7456$ & $0.7445$ & $0.0038$ & $0.7342$ & $0.7340$ 
    & $0.0040$ & $0.7356$ & $0.7373$ & $0.0052$\\ \hline
    $H_0$               & $71.624$ & $71.521$ & $0.3556$ & $71.089$ & $71.075$ 
    & $0.3570$ & $71.265$ & $71.430$ & $0.4931$\\ \hline
    $z_{\rm re}$        & $10.126$ & $10.132$ & $0.2250$ & $10.443$ & $10.366$ 
    & $0.2285$ & $10.397$ & $10.448$ & $0.2319$\\ \hline
  \end{tabular}            
  \caption{Best fits, mean values and standard deviations for cosmological
    parameters when $B$ modes are included in the $ C_\ell$-likelihood. All values
    are rounded to order $10^{-4}$ to the nearest value and correspond
to the cumulative channel whose $ \chi^2 $ is the
  sum of the $ \chi^2 $'s of the three channels HFI-100, HFI-143 and LFI-70.}
  \label{tab4}
  \end{table}

\begin{table}
  \centering
   \begin{tabular}{|p{1.8cm}|p{2cm}||p{1.8cm}|p{1.8cm}||p{1.8cm}|p{1.8cm}||p{1.8cm}|p{1.8cm}|}
%  \begin{tabular}{|c|c||c|c||c|c||c|c|}
    \hline
    \multirow{3}{*}{}  & sky data& \multicolumn{2}{c||}{$\Lambda$CDM: fiducial $ r=0 $} &
    \multicolumn{4}{c|}{$\Lambda$CDM$r$T: fiducial $ r=0.0427 $}  \\  \cline{2-8}
    & test model&\multicolumn{2}{c||}{$\Lambda$CDM$r$}
    &\multicolumn{2}{c||}{$\Lambda$CDM$r$ }
    &\multicolumn{2}{c|}{$\Lambda$CDM$r$T }     \\  \cline{2-8}
    & & $68$\% CL & $95$\% CL & $68$\% CL & $95$\% CL & $68$\% CL & $95$\% CL \\
    \hline
    \multirow{4}*{\minitab[c]{without\\ $B$ modes}}
    & LFI--70 & $r<0.2$ & $r<0.2$ & $r<0.2$ & $r<0.2$
    & $\,r<0.102$ & $\,r<0.128$ \\
    \cline{2-8}
    & HFI--100 & $\,r<0.068 $ & $\,r<0.124$ & $\,r<0.097$ & $r<0.155$
    & $r<0.047$ & $r<0.111$ \\
    \cline{2-8}
    & HFI--143 & $r<0.070$ & $r<0.117$ & $r<0.108$ &$r<0.158$
    & $r<0.042$ & $r<0.061$ \\
    \cline{2-8}
    & Cumulative & $r<0.036$ & $r<0.068$ & $r<0.066$ &$r<0.102$
    & $r<0.051$ & $r<0.113$ \\
    \hline
    \multirow{4}*{\minitab[c]{with\\ $B$ modes}}
    & LFI--70 & $r<0.074$ & $r<0.151$ & $r<0.075$ & $r<0.144$
    & $r<0.117$ & $r<0.131$ \\
    \cline{2-8}
    & HFI--100 & $r<0.012$ & $r<0.029$ & $r<0.037$
    & $r<0.065$ & $r<0.049$ & $r<0.112$ \\
    \cline{2-8}
    & HFI--143 & $r<0.008$ & $r<0.020$ & $r<0.041\,$
    & $r<0.064\,$ & $r<0.042$ & $r<0.062$ \\
    \cline{2-8}
    & Cumulative & $r<0.008$ & $r<0.016$ &$r<0.032$
    & $r<0.045$ & $r<0.052$ & $r<0.114$ \\
    \hline
  \end{tabular}
  \caption{Upper bounds on $ r $ with all figures rounded upward to order
    $10^{-3}$. Notice that the bound $r<0.2$ is just the assumed prior, which
    gets saturated by the $\Lambda$CDM$r$ test model in the LFI--70 channel when
    $ B $ modes are absent. The limits in the case of the $\Lambda$CDM$r$T test
    model with fiducial $r=0.0427$ in the LFI--70 channel are not really significant
    in view of the shape of the corresponding likelihood distribution (see
    Fig. \ref{r04br04T}).}
      \label{tab5}
      \end{table}

\begin{table}
  \centering
  \begin{tabular}{|c|c||c|c||c|c|}
    \hline
    \multirow{3}{*}{}  & sky data& 
    \multicolumn{4}{c|}{$\Lambda$CDM$r$T: fiducial $ r=0.0427 $}  \\  \cline{2-6}
    & test model    &\multicolumn{2}{c||}{$\Lambda$CDM$r$ }
    &\multicolumn{2}{c|}{$\Lambda$CDM$r$T }     \\  \cline{2-6}
    & & $68$\% CL & $95$\% CL & $68$\% CL & $95$\% CL \\
    \hline
    \multirow{4}*{\minitab[c]{without\\ $B$ modes}}
    & LFI--70  & & & $\,r>0.067$ & $r>\,0.035$ \\
    \cline{2-6}
    & HFI--100  & $\,r>0.051$ & & $\,r>0.034$ & $\,r>0.024$ \\
    \cline{2-6}
    & HFI--143  & $r>0.034$ &$\,r>0.026$& $\,r>0.030$ & $\,r>0.025$ \\
    \cline{2-6}
    & Cumulative  & $r>0.024$ & & $\,r>0.034$ & $\,r>0.030$ \\
    \hline
    \multirow{4}*{\minitab[c]{with\\ $B$ modes}}
    & LFI--70  & & & $\,r>0.046$ & $\,r>0.024$ \\
    \cline{2-6}
    & HFI--100  & $\,r>0.020$ & & $\,r>0.034$ & $\,r>0.025$ \\
    \cline{2-6}
    & HFI--143  & $\,r>0.026$ & $\,r>0.013$ & $\,r>0.033$ & $\,r>0.026$ \\
    \cline{2-6}
    & Cumulative  & $r>0.022$ & $r>0.013$ & $\,r>0.039$ & $\,r>0.030$ \\
    \hline
  \end{tabular}
  \caption{Lower bounds on $ r $ with all figures rounded downward to order
    $10^{-3}$. These bounds are assumed significant only when
    the likelihood at $r=0$ is less than $ \exp(-1/2)=0.6065\ldots $ of its maximum
    for $68$\% CL or less than $ \exp(-1)=0.3678\ldots $ for $95$\% CL.
    The limits in the case of the $\Lambda$CDM$r$T test
    model with fiducial $ r=0.0427 $ in the LFI--70 channel are not really significant
    in view of the shape of the corresponding likelihood distribution as can be seen from
    Fig. \ref{r04br04T}. In the $\Lambda$CDM$r$ model, the entries left empty in the table
correspond to the cases where there are no lower bounds on $ r $ (as can be seen from Figs. 
\ref{r0r04} and \ref{br0br04}).}
  \label{tab6}
  \end{table}

\begin{table}[h]
  \centering
  \begin{tabular}{|p{2.5cm}||p{1.2cm}|p{1.2cm}|p{1.2cm}||p{1.2cm}|p{1.2cm}|p{1.2cm}||p{1.2cm}|p{1.2cm}|p{1.2cm}|}
%  \begin{tabular}{|c||c|c|c||c|c|c||c|c|c|}
    \hline
    sky data& \multicolumn{3}{c||}{$\Lambda$CDM: fiducial $ r=0 $} &
    \multicolumn{6}{c|}{$\Lambda$CDM$r$T: fiducial $ r=0.0427 $}  \\  
    \hline
    test model&\multicolumn{3}{c||}{$\Lambda$CDM$r$} &
    \multicolumn{3}{c||}{$\Lambda$CDM$r$ } &
    \multicolumn{3}{c|}{$\Lambda$CDM$r$T }     \\
    \hline 
    &~~best~~ & ~~mean~~ & stddev~ 
    & ~~best~~ & ~~mean~~ & stddev~ 
    & ~~best~~ & ~~mean~~ & stddev~\\
    \hline
    no residuals & $0.0010$ & $0.0060$ & $0.0050$ 
                 & $0.0240$ & $0.0275$ & $0.0096$ 
                 & $0.0405$ & $0.0516$ & $0.0230$\\
    \hline
    best case smooth & $0.0040$ & $0.0222$ & $0.0160$ 
                     & $0.0448$ & $0.0504$ & $0.0238$ 
                     & $0.0465$ & $0.0516$ & $0.0235$\\
    \hline
    middle case rugged~& $0.0024$ & $0.0230$ & $0.0188$ 
                        & $0.0431$ & $0.0472$ & $0.0261$ 
                        & $0.0344$ & $0.0513$ & $0.0234$\\ 
    \hline
    worst case smooth & $0.0083$ & $0.0250$ & $0.0160$ 
                      & $0.0436$ & $0.0480$ & $0.0275$ 
                      & $0.0387$ & $0.0518$ & $0.0250$\\
    \hline
  \end{tabular}
  \caption{Best fits, mean values and standard deviations of the ratio $r$
    when $B$ modes are included in the cumulative $ C_\ell-$likelihood and
    foreground residuals are taken into account. All figures
    are rounded to order $10^{-4}$ to the nearest value.}
      \label{tab7}
      \end{table}

\begin{table}[h]
  \centering
  \begin{tabular}{|c||c|c||c|c||c|c|}
    \hline
    sky data& \multicolumn{2}{c||}{$\Lambda$CDM: fiducial $ r=0 $} &
    \multicolumn{4}{c|}{$\Lambda$CDM$r$T: fiducial $ r=0.0427 $}  \\  
    \hline
    test model&\multicolumn{2}{c||}{$\Lambda$CDM$r$} &
    \multicolumn{2}{c||}{$\Lambda$CDM$r$ } &
    \multicolumn{2}{c|}{$\Lambda$CDM$r$T }     \\
    \hline 
&$68$\% CL & $95$\% CL & $68$\% CL & $95$\% CL & $68$\% CL & $95$\% CL \\
    \hline
    no residuals & $\,r<0.008$ & $\,r<0.016$ &$\,r<0.032$
    & $\,r<0.045$ & $\,r<0.052$ & $\,r<0.114$ \\
    \hline
    best case smooth & $r<0.028$ & $r<0.053$ & $r<0.062$
    & $r<0.091$ & $r<0.052$ & $r<0.115$ \\
    \hline
    ~middle case rugged~ & $r<0.029$ & $r<0.058$ & $r<0.059\,$
    & $\,r<0.094\,$ & $r<0.052$ & $r<0.115$ \\
    \hline
    worst case smooth & $r<0.032$ & $r<0.062$ &$r<0.060$ 
    & $\,r<0.097$ & $r<0.052$ & $r<0.116$ \\
    \hline
  \end{tabular}
  \caption{Upper bounds on $ r $ when foreground residuals are
    considered, using the cumulative likelihoods and including $B$ modes.
    All figures are rounded upward to order $10^{-3}$.}
      \label{tab8}
      \end{table}

\begin{table}[h]
  \centering
  \begin{tabular}{|c||c|c||c|c||c|c|}
    \hline
    sky data& \multicolumn{2}{c||}{$\Lambda$CDM: fiducial $ r=0 $} &
    \multicolumn{4}{c|}{$\Lambda$CDM$r$T: fiducial $ r=0.0427 $}  \\  
    \hline
    test model&\multicolumn{2}{c||}{$\Lambda$CDM$r$} &
    \multicolumn{2}{c||}{$\Lambda$CDM$r$ } &
    \multicolumn{2}{c|}{$\Lambda$CDM$r$T }     \\
    \hline 
&~~$68$\% CL~~& $95$\% CL & $68$\% CL & $95$\% CL & $68$\% CL & 
$95$\% CL \\
    \hline
no residuals  & & & $\,r>0.022$ & $\,r>0.013$ & $\,r>0.039$ & 
$\,r>0.030$ \\
    \hline
 best case smooth & & & $r>0.037$ & $r>0.012$ & $r>0.038$ & $r>0.029$ \\
    \hline
    ~middle case rugged~ & & & $r>0.032$ & & $r>0.038$ & $r>0.029$ \\
    \hline
    worst case smooth & & &$r>0.032$ & & $r>0.037$ & $r>0.028$ \\
    \hline
  \end{tabular}
  \caption{Lower bounds on $ r $ when foreground residuals are
    considered, using the cumulative likelihoods and including $B$ modes.
    These bounds are assumed significant only when
 the likelihood at $ r=0 $ is less than $ \exp(-1/2)=0.6065\ldots $ of its 
maximum for $68$\% CL or less than $ \exp(-1)=0.3678\ldots $ for $95$\% CL. 
Otherwise the entry is left empty in this table.
All figures are rounded downward to order $ 10^{-3} $.}
      \label{tab9}
\end{table}      

%FIGURES%
\clearpage

\end{document}